%% file: main.tex
\documentclass[superscriptaddress,twocolumn,nofootinbib,floatfix]{revtex4-1}
\usepackage{amssymb,amsmath,amsfonts,amsthm,dsfont}
\usepackage{bbm}
\usepackage{physics,gensymb}
\usepackage{graphicx,wasysym}
\usepackage{float}
\usepackage[normalem]{ulem}
\usepackage[lofdepth,lotdepth,caption=false]{subfig}
\usepackage{color}
\usepackage{listings}
\usepackage{rotating}
\usepackage{qcircuit}
\usepackage{url}

\newcommand{\RR}{\mathbbm{R}}
\newcommand{\x}{\mathbf{x}}

\begin{document}

\setlength{\parskip}{0pt}

\title{Quantum classifier with tailored quantum kernel}

\author{Carsten Blank}
\email{blank@data-cybernetics.com }
\affiliation{Data Cybernetics, 86899 Landsberg, Germany}
\author{Daniel K. Park}
\email{dkp.quantum@gmail.com}
\affiliation{School of Electrical Engineering, KAIST, Daejeon, 34141, Republic of Korea}
\affiliation{ITRC of Quantum Computing for AI, KAIST, Daejeon, 34141, Republic of Korea}
\author{June-Koo Kevin Rhee}
\email{rhee.jk@kaist.edu}
\affiliation{School of Electrical Engineering, KAIST, Daejeon, 34141, Republic of Korea}
\affiliation{ITRC of Quantum Computing for AI, KAIST, Daejeon, 34141, Republic of Korea}
\affiliation{Quantum Research Group, School of Chemistry and Physics, University of KwaZulu-Natal, Durban, KwaZulu-Natal, 4001, South Africa}
\author{Francesco Petruccione}
\email{petruccione@ukzn.ac.za}
\affiliation{School of Electrical Engineering, KAIST, Daejeon, 34141, Republic of Korea}
\affiliation{Quantum Research Group, School of Chemistry and Physics, University of KwaZulu-Natal, Durban, KwaZulu-Natal, 4001, South Africa}
\affiliation{National Institute for Theoretical Physics (NITheP), KwaZulu-Natal, 4001, South Africa}
\begin{abstract}
Kernel methods have a wide spectrum of applications in machine learning. Recently, a link between quantum computing and kernel theory has been formally established, opening up opportunities for quantum techniques to enhance various existing machine learning methods. We present a distance-based quantum classifier whose kernel is based on the quantum state fidelity between training and test data. The quantum kernel can be tailored systematically with a quantum circuit to raise the kernel to an arbitrary power and to assign arbitrary weights to each training data. Given a specific input state, our protocol calculates the weighted power sum of fidelities of quantum data in quantum parallel via a swap-test circuit followed by two single-qubit measurements, requiring only a constant number of repetitions regardless of the number of data. We also show that our classifier is equivalent to measuring the expectation value of a Helstrom operator, from which the well-known optimal quantum state discrimination can be derived. We demonstrate the performance of our classifier via classical simulations with a realistic noise model and proof-of-principle experiments using the IBM quantum cloud platform.
\end{abstract}
\maketitle
\def\one{{\mathchoice {\rm 1\mskip-4mu l} {\rm 1\mskip-4mu l} {\rm \mskip-4.5mu l} {\rm 1\mskip-5mu l}}}
\section*{Introduction}
Advances in quantum information science and machine learning have led to the natural emergence of quantum machine learning, a field that bridges the two, aiming to revolutionize information technology~\cite{wittek,doi:10.1080/00107514.2014.964942,QML-Biamonte,SupervisedQML,Dunjko_2018}. The core of its interest lies in either taking advantage of quantum effects to achieve machine learning that surpasses the classical pendant in terms of computational complexity or to entirely be able to apply such techniques on quantum data. A prominent application of machine learning is classification for predicting a category of an input data by learning from labeled data, an example of pattern recognition in big data analysis. As most techniques in classical supervised machine learning are aimed to getting the best result while using a polynomial amount of computational resources at most, an exact solution to the problem is usually out of reach. Therefore many such learning protocols have empirical scores instead of analytically calculated bounds. Even with this lack of rigorous mathematics they have been applied with great success in science and industry. In pattern analysis, the use of a kernel, i.e. a similarity measure of data that corresponds to an inner product in higher-dimensional feature space, is vital~\cite{Scholkopf:2000:KTD:3008751.3008793,hofmann2008}. However, classical classifiers that rely on kernel methods are limited when the feature space is large and the kernel functions are computationally expensive to evaluate. Recently, a link between the kernel method with feature maps and quantum computation was formally established by proposing to use quantum Hilbert spaces as feature spaces for data~\cite{PhysRevLett.122.040504}. The ability of a quantum computer to efficiently access and manipulate data in the quantum feature space offers potential quantum speedups in machine learning~\cite{Havlicek2019}.

Recent work in Ref.~\cite{QML_Maria_Francesco} showed a minimal quantum interference circuit for realizing a distance-based supervised binary classifier. The goal of this task is, given a labelled data set $\mathcal{D} = \left\{ (\x_1, y_1), \ldots, (\x_M, y_M) \right\} \subset \mathbb{C}^N\times\{0,1\}$, to classify an unseen data point $\tilde{\x} \in \mathbb{C}^N$ as best as possible. Conventional machine learning problems usually deal with real-valued data points, which is however not the natural choice for quantum information problems. In particular, having quantum feature maps in mind, we generalize the data set to be complex valued. The quantum interference circuit introduced in Ref.~\cite{QML_Maria_Francesco} implements a distance-based classifier through a kernel based on the real part of the transition probability amplitude (state overlap) between training and test data. Once the set of classical data is encoded as a quantum state in a specific format, the classifier can be implemented by interfering the training and test data via a Hadamard gate and gathering the projective measurement statistics on a post-selected state which has been projected to a particular subspace. For brevity, we refer to this classifier as \textit{Hadamard classifier}. Since a Hadamard classifier only takes the real part of the state overlap into account it does not work for an arbitrary quantum state, which can represent classical data via a quantum feature map or be an intrinsic quantum data. Thus, designing quantum classifiers that work for an arbitrary quantum state is of fundamental importance for further developments of quantum methods for supervised learning.

In this work, we propose a distance-based quantum classifier whose kernel is based on the quantum state fidelity, thereby enabling the use of a quantum feature map to the full extent. We present a simple and systematic construction of a quantum circuit for realizing an arbitrary weighted power sum of quantum state fidelities between the training and test data as the distance measure. The argument for the introduction of non-uniform weights can also be applied to the Hadamard classifier of Ref.~\cite{QML_Maria_Francesco}. The classifier is realized by applying a swap-test~\cite{PhysRevLett.87.167902} to a quantum state that encodes the training and test data in a specific format. The quantum state fidelity can be raised to the power of $n$ at the cost of using $n$ copies of training and test data. We also show that the post-selection can be avoided by measuring an expectation value of a two-qubit observable. The swap-test classifier can be implemented without relying on the specific initial state by using a method based on \textit{quantum forking}~\cite{ffqram,qfs} at the cost of increasing the number of qubits. In this case, the training data, corresponding labels, and the test data are provided on separate registers as a product state. This approach is especially useful for a number of situations: intrinsic---possibly unknown---quantum data, parallel state preparation and gate intensive routines, such as quantum feature maps. Furthermore, we show that the swap-test classifier is equivalent to measuring the expectation value of a Helstrom operator, from which the optimal projectors for the quantum state discrimination is constructed~\cite{Helstrom1969}. This motivates further investigations on the fundamental connection between the distance-based quantum classification and the Helstrom measurement. To demonstrate the feasibility of the classifier with near-term quantum devices, we perform simulations on a classical computer with a realistic error model, and realize a proof-of-principle experiment on a five-qubit quantum computer in the cloud provided by IBM~\cite{ibm_q_experience}.
\section*{Results}
\subsection*{Classification without post-selection}
The Hadamard classifier requires the training and test data to be prepared in a quantum state as
		\begin{equation}
			\label{qtml:state}
			\ket{\Psi^h} = \frac{1}{\sqrt{2}} \sum_{m=1}^{M} \sqrt{w_m}\left(\ket{0}\ket{\x_m} + \ket{1}  \ket{\tilde{\x}} \right)\ket{y_m}\ket{m} ,
		\end{equation}
where the data are encoded into the state representation $\ket{\x_m} = \sum_{i=1}^{N} x_{m,i} \ket{i}$, $\ket{\tilde{\x}} = \sum_{i=1}^{N} \tilde{x}_{i} \ket{i}$, the binary label is encoded in $y_m\in\lbrace 0,1 \rbrace$, and all inputs $\x_m$ and $\tilde{\x}$ have unit length~\cite{QML_Maria_Francesco}. The superscript $h$ indicates that the state is for the Hadamard classifier. The first and the last qubits are an ancilla qubit used for interfering training and test data and index qubits for training data, respectively. In Ref.~\cite{QML_Maria_Francesco}, each subspace has an equal probability amplitude, i.e., $w_m=1/M\;\forall\; m$, resulting in a uniformly weighted kernel. Here we introduce an arbitrary probability amplitude $\sqrt{w_m}$, where $\sum_m w_m=1$, to show that a non-uniformly weighted kernel can also be generated. The goal of the classifier is to assign a new label $\tilde{y}$ to the test data, which predicts the true class of $\tilde{\x}$ denoted by $c(\tilde{\x})$ with high probability. The classifier is implemented by a quantum interference circuit consisting of a Hadamard gate and two single-qubit measurements. The state after the Hadamard gate applied to the ancilla qubit is
	\begin{equation}\label{eqn:after_hadamard}
	H\ket{\Psi^h} =\frac{1}{2} \sum_{m=1}^{M} \sqrt{w_m}\left( \ket{0}  \ket{\psi_{+}} + \ket{1} \ket{\psi_{-}} \right)\ket{y_m}\ket{m}
	\end{equation}
with $\ket{\psi_\pm} = \ket{\x_m} \pm \ket{\tilde{\x}}$. Measuring the ancilla qubit in the computational basis and post-selecting the state $\ket{a}$, $a\in\lbrace0,1\rbrace$, yield the state
	\begin{equation}\label{eqn:classifier_state}
	\ket{\Psi_a^h} = \frac{1}{2\sqrt{p_{a}}} \sum_{m=1}^{M}\sqrt{w_m} \ket{a} \ket{\psi_a}\ket{y_m}\ket{m},
	\end{equation}
where $p_{a}=\sum_{m=1}^Mw_m (1+(-1)^a\text{Re}\braket{\psi_{\x_m}}{\psi_{\tilde{\x}}})/2$ is the probability to post-select $a=0$ or 1, and $\psi_{0(1)}=\psi_{+(-)}$. The Hadamard classifier in Ref.~\cite{QML_Maria_Francesco} selects the measurement outcome $a=0$ and proceeds with a measurement of the label register in the computational basis, resulting in the measurement probability of
	\begin{equation}
	    \label{eqn:kernel+}
	    \begin{aligned}
		\mathbbm{P}(\tilde{y} = b|a=0) =&\text{tr}\left\lbrack \left(\one^{\otimes 2}\otimes \ketbra{b}{b}\otimes\one\right)\ketbra{\Psi_0^h}{\Psi_0^h}\right\rbrack \\
	        =& \frac{1}{2p_{0}} \sum_{m |y_m = b}^M w_m\left(1 + \text{Re}\braket{\tilde{\x}}{\x_m}\right),
	    \end{aligned}
	\end{equation}
where $b\in\lbrace 0,1 \rbrace$. The test data is classified as $\tilde y$ that is obtained with a higher probability. Since the success probability of the classification depends on $p_0$, in Ref.~\cite{QML_Maria_Francesco}, a data set is to be pre-processed in a way that the post-selection succeeds with a probability of around $1/2$. This is done by standardizing all data $\x_m$ such that they have mean $0$ and standard deviation $1$ and applying the transformation to the test datum $\tilde{\x}$ too. Now we show that the classifier can be realized without the post-selection, thereby reducing the number of experiments by about a factor of two, and avoiding the pre-processing (see Supplementary Information). 
	
If the classifier protocol proceeds with the ancilla qubit measurement outcome of $1$, the probability to measure $b$ on the label qubit is
	\begin{equation}
	    \label{eqn:kernel-}
	    \begin{aligned}
		\mathbbm{P}(\tilde{y} = b|a=1) =&\text{tr}\left\lbrack \left(\one^{\otimes 2}\otimes \ketbra{b}{b}\otimes\one\right)\ketbra{\Psi_1^h}{\Psi_1^h}\right\rbrack \\
	        =& \frac{1}{2p_{1}} \sum_{m|y_m = b}^M w_m\left( 1 - \text{Re}\braket{\tilde{\x}}{\x_m}\right).
	    \end{aligned}
	\end{equation}
Thus, when the ancilla qubit measurement outputs 1, $\tilde y$ should be assigned to the label with a lower probability. This result shows that both branches of the ancilla state can be used for classification. The difference in the post-selected branch only results in different post-processing of the measurement outcomes.

The measurement and the post-processing procedure can be described more succinctly with an expectation value of a two-qubit observable, $\langle \sigma_z^{(a)}\sigma_z^{(l)}\rangle$, where the superscript $a$ ($l$) indicates that the operator is acting on the ancilla (label) qubit. The expectation value is

\begin{align}
	\label{eqn:hadamard_kernel}
    &\langle \sigma_z^{(a)}\sigma_z^{(l)}\rangle=\text{tr}\left(\sigma_z^{(a)}\sigma_z^{(l)}H\ketbra{\Psi^h}{\Psi^h}H\right)\nonumber \\
    &=\sum_{m=1}^M\frac{w_m}{4}\Big{[}\text{tr}\left(\sigma_z\!\ketbra{0}{0}\otimes\ketbra{\psi_+}{\psi_+}\otimes\sigma_z\!\ketbra{y_m}{y_m}\right)\nonumber \\
    &\qquad\qquad\;+\text{tr}\left(\sigma_z\!\ketbra{1}{1}\otimes\ketbra{\psi_-}{\psi_-}\otimes\sigma_z\!\ketbra{y_m}{y_m}\right)\Big{]}\nonumber \\
    &=\sum_{m=1}^M\frac{w_m}{4}\left\lbrack\text{tr}\left(\ketbra{\psi_+}{\psi_+}\right)\!-\!\text{tr}\left(\ketbra{\psi_-}{\psi_-}\right)\right\rbrack\text{tr}\left(\sigma_z\!\ketbra{y_m}{y_m}\right)\nonumber \\
    &=\sum_{m=1}^M(-1)^{y_m}w_m\text{Re}\braket{\tilde{\x}}{\x_m}.
\end{align}
The last expression is obtained by using $\text{tr}(\ketbra{\psi_\pm}{\psi_\pm})=2\pm 2\text{Re}\braket{\tilde{\x}}{\x_m}$, and $\text{tr}(\sigma_z\ketbra{y_m}{y_m})=1$ for $y_m=0$ and $-1$ for $y_m=1$. The test data is classified as 0 if $\langle \sigma_z^{(a)}\sigma_z^{(l)}\rangle$ is positive, and 1 if negative:

	\begin{equation}
	\label{eq:y-tilde}
	\tilde{y}=\frac{1}{2}\left(1-\text{sgn}\left(\langle\sigma_z^{(a)}\sigma_z^{(l)}\rangle\right)\right).
	\end{equation}
	
	\begin{figure}[t]
        \centering
        \includegraphics[width=0.8\columnwidth]{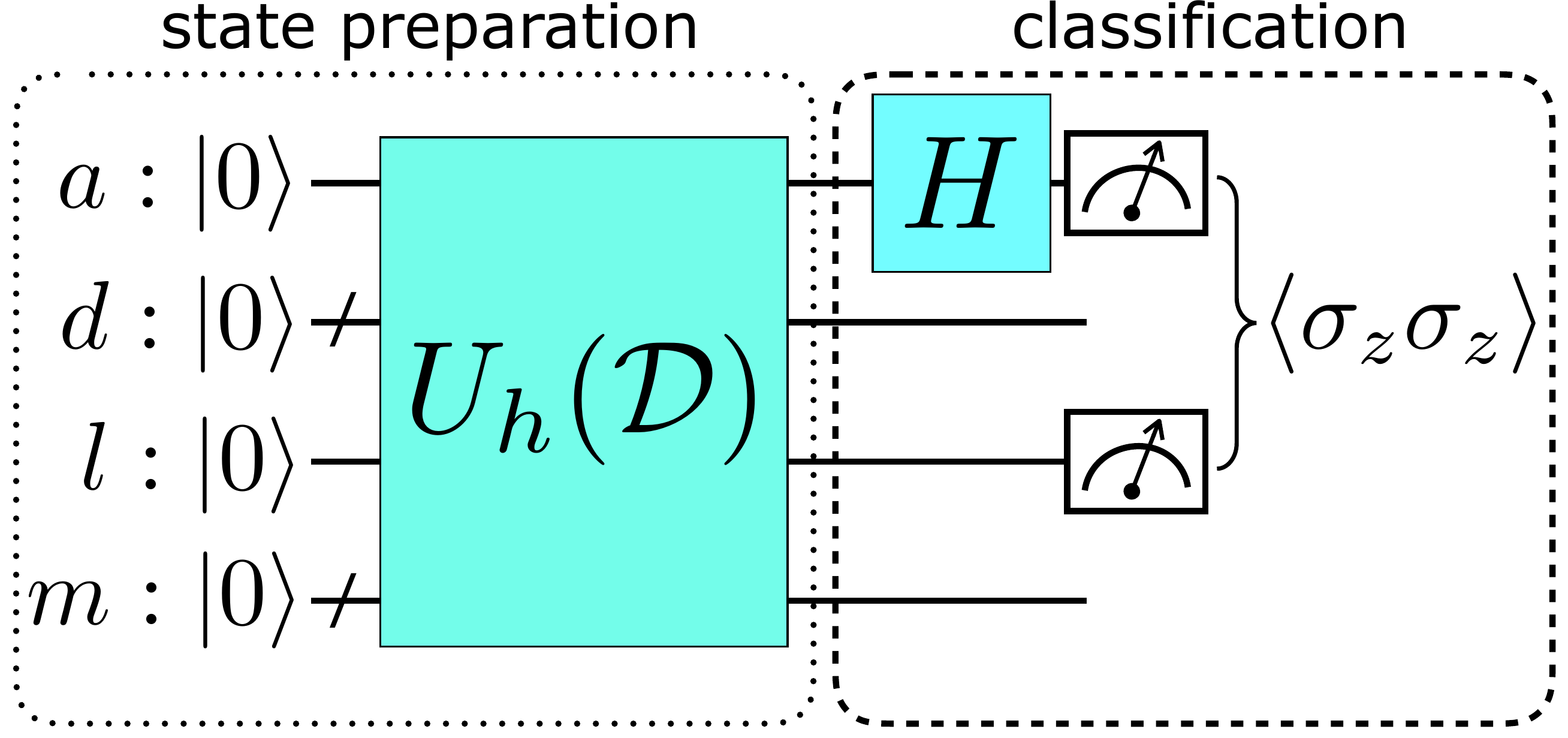}
        \caption{The Hadamard classifier. The first register is the ancilla qubit ($a$), the second is the data qubit ($d$), third is the label qubit ($l$), and the last one corresponds to the index qubits ($m$). An operator $U_h(\mathcal{D})$ creates the input state necessary for the classification protocol. The Hadamard gate and the two-qubit measurement statistics yield the classification outcome.\label{fig:HadamardClassifier}}
	\end{figure}
\subsection*{Quantum kernel based on state fidelity}
In order to take the full advantage of the quantum feature maps~\cite{PhysRevLett.122.040504,Havlicek2019} in the full range of machine learning applications, it is desirable to construct a kernel based on the quantum state fidelity, rather than considering only a real part of the quantum state overlap as done in Ref.~\cite{QML_Maria_Francesco}. We propose a quantum classifier based on the quantum state fidelity by using a different initial state than described in Ref.~\cite{QML_Maria_Francesco} and replacing the Hadamard classification with a swap-test.
    
The state preparation requires the training data with labels to be encoded as a specific format in the index, data and label registers. In parallel, a state preparation of the test data is done on a separate input register. Unlike in the Hadamard classifier, the ancilla qubit is not in the part of the state preparation, and it is only used in the measurement step as the control qubit for the swap-test. The controlled-swap gate exchanges the training data and the test data, and the classification is completed with the expectation value measurement of a two-qubit observable on the ancilla and the label qubits. For brevity, we refer to this classifier as \textit{swap-test classifier}.

\begin{figure}[t]
        \centering
        \includegraphics[width=0.98\columnwidth]{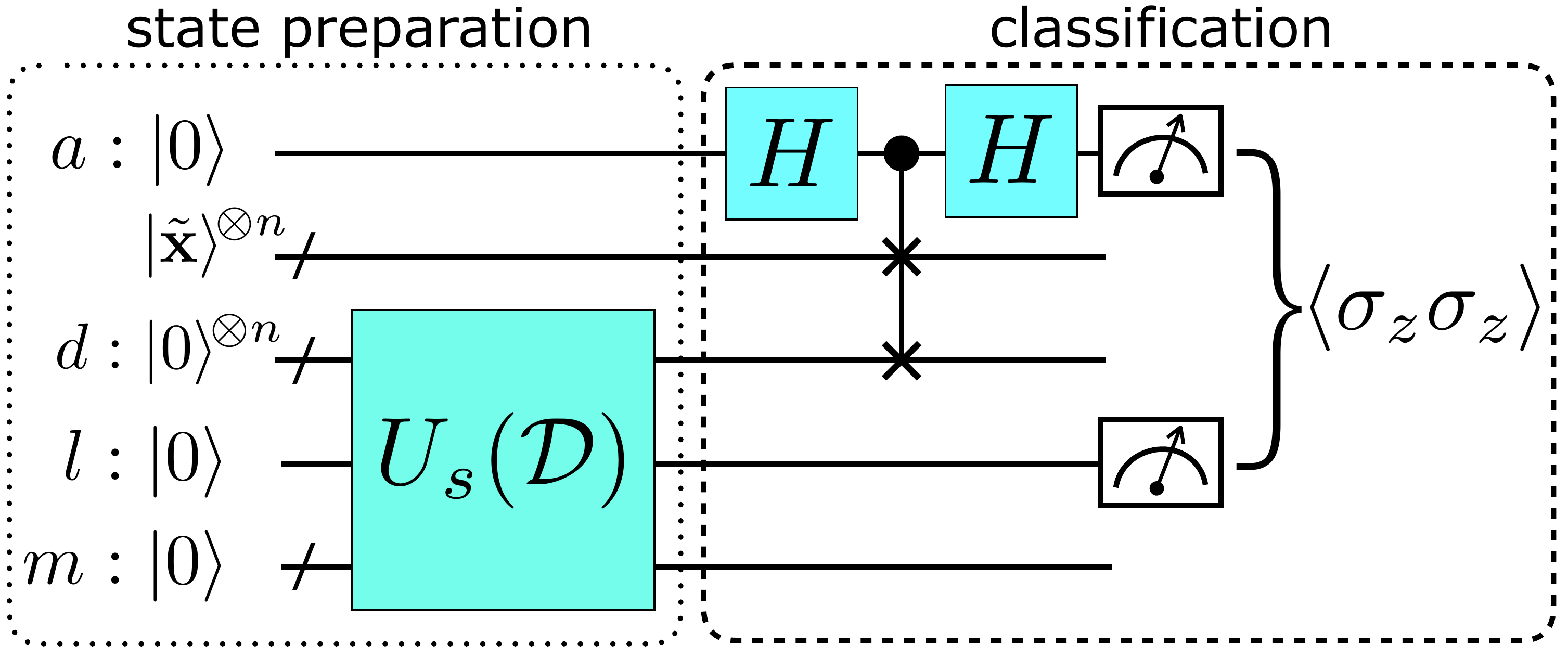}
        \caption{The swap-test classifier. The first register is the ancilla qubit ($a$), the second contains $n$ copies of the test datum ($\tilde{x}$), the third are the data qubits ($d$), the fourth is the label qubit ($l$) and the final register corresponds to the index qubits ($m$). An operator $U_s(\mathcal{D})$ creates the input state necessary for the classification protocol. The swap-test and the two-qubit measurement statistics yield the classification outcome. \label{fig:SwapTestClassifier}}
    \end{figure}

With multiple copies of training and test data, polynomial kernels can be designed~\cite{PhysRevLett.113.130503,cagliari.hqc}. With any $n \in \mathbb{N}$, a swap-test on $n$ copies of training and test data that are entangled in a specific form results in

\begin{align}
    \label{eq:general_state}
     &\sum_{m=1}^M\sqrt{w_m}\ket{0}\ket{\tilde{\x}}^{\otimes n}\ket{\x_m}^{\otimes n}\ket{y_m}\ket{m}\xrightarrow{H_a\cdot\text{c-}\texttt{swap}^n\cdot H_a}\ket{\Psi^s_f}\nonumber\\
    &=\sum_{m=1}^M\frac{\sqrt{w_m}}{2}(\ket{0}\ket{\psi_{n+}}+\ket{1}\ket{\psi_{n-}})\ket{y_m}\ket{m},
\end{align}
where $\ket{\psi_{n\pm}}=\ket{\tilde{\x}}^{\otimes n}\ket{\x_m}^{\otimes n}\pm\ket{\x_m}^{\otimes n}\ket{\tilde{\x}}^{\otimes n}$, and the superscript $s$ indicates that the state is for the swap-test classifier. Using $\text{tr}(\ketbra{\psi_{n\pm}}{\psi_{n\pm}})=2\pm 2|\braket{\tilde\x}{\x_m}|^{2n}$, the expectation value of $\sigma_z^{(a)}\sigma_z^{(l)}$ for this state is given as
\begin{equation}
    \label{eq:DesignedKernel}
        \text{tr}\left(\sigma_z^{(a)}\sigma_z^{(l)}\ketbra{\Psi_f^s}{\Psi_f^s}\right)
		=\sum_{m=1}^M(-1)^{y_m}w_m|\braket{\tilde\x}{\x_m}|^{2n}.
    \end{equation}
The swap-test classifier also assigns a label to the test data according to Eq.~(\ref{eq:y-tilde}). A quantum circuit for implementing a swap-test classifier with a kernel based on the $n$th power of the quantum state fidelity is depicted in Fig.~\ref{fig:SwapTestClassifier}.

Note that if the projective measurement in the computational basis followed by post-selection is performed as in Ref.~\cite{QML_Maria_Francesco}, the probability of classification can be obtained as
    \begin{equation}
    \label{eq:DesignedKernelPostSelect}
        \mathbbm{P}(\tilde{y} = b|a) =\frac{1}{2p_{a}} \sum_{m |y_m = b}^M w_m\left( 1 + (-1)^a |\braket{\tilde{\x}}{\x_m}|^{2n}\right),
    \end{equation}
where $p_{a}=\sum_m^M w_m(1 + (-1)^a |\braket{\tilde{\x}}{\x_m}|^{2n})/2$. Since $p_a$ here is a function of the quantum state fidelity, which is non-negative, $p_0\ge p_1$ and $p_0\ge 1/2$. As a result, the data pre-processing used in the Hadamard classifier for ensuring a high success probability of the post-selection is not strictly required for the swap-test classifier.

\begin{figure}[t]
		\centering
		\includegraphics[width=1\columnwidth]{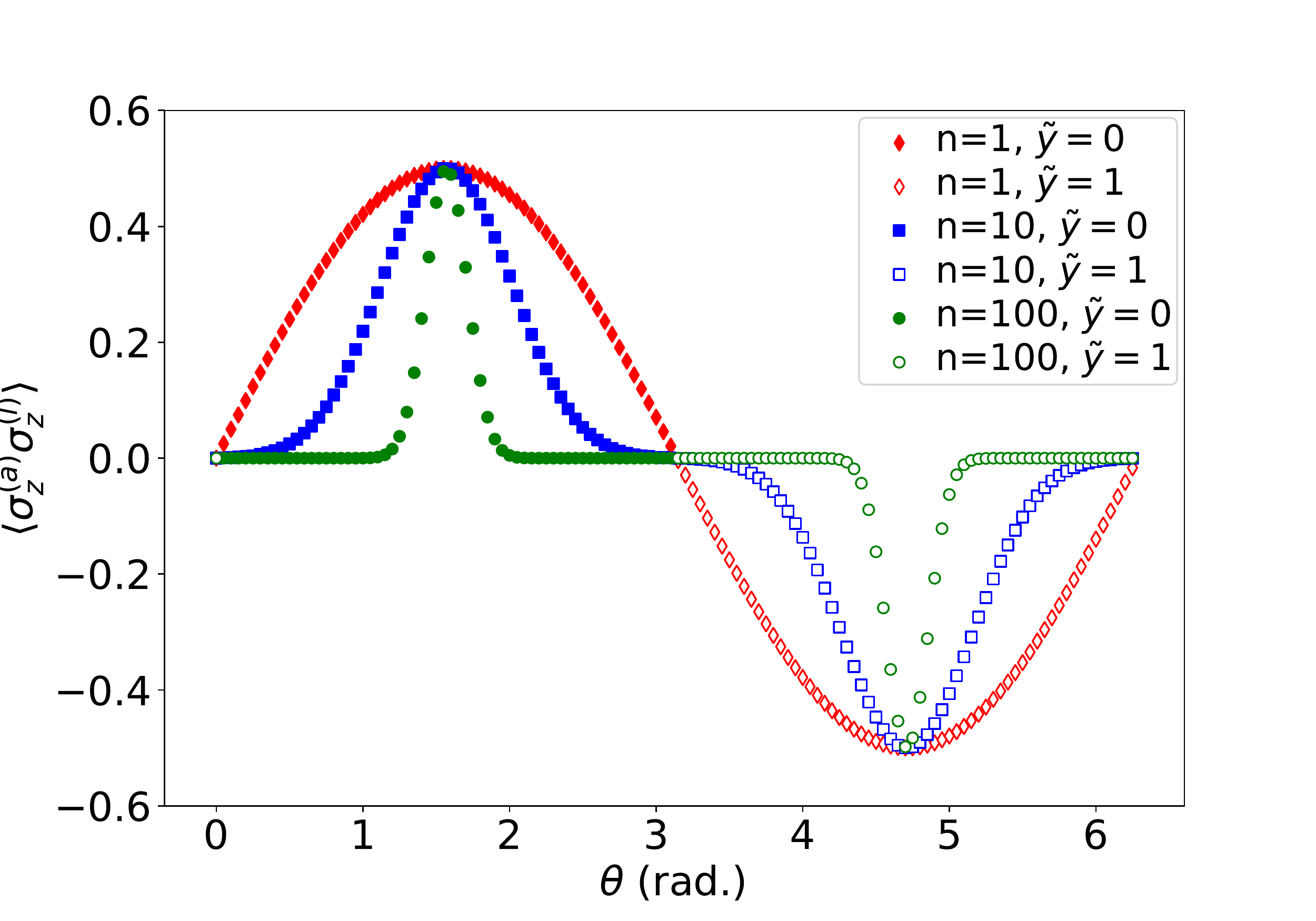}
		\caption{Theoretical results of the swap-test classifier for the example given in Eq.~(\ref{eqn:data_example}), for $n=$ 1, 10, and 100 copies of training and test data. The test data is classified as 0 (1) if the expectation value, $\langle \sigma_z^{(a)} \sigma_z^{(l)} \rangle$, is positive (negative). The comparison of the results for various $n$ illustrates the polynomial sharpening which will eventually result into a Dirac $\delta$ if the number of copies approaches to the limit of $\infty$. \label{fig:classifier_experiment_copies}}
	\end{figure}

We demonstrate the performance of the swap-test classifier using a simple example data set that only consists of two training data and one test data as

	\begin{align}
		\label{eqn:data_example}
		&\ket{\x_1} = \frac{i}{\sqrt{2}} \ket{0} + \frac{1}{\sqrt{2}} \ket{1},\; y_1 = c(\x_1) = 0,\nonumber\\
		&\ket{\x_2} = \frac{i}{\sqrt{2}} \ket{0} - \frac{1}{\sqrt{2}} \ket{1},\; y_2 = c(\x_2) = 1,\nonumber \\
		&\ket{\tilde{\x}(\theta)} = \cos{\frac{\theta}{2}} \ket{0} - i \sin{\frac{\theta}{2}} \ket{1}, \nonumber \\
		&c(\x) = \frac{1}{2} \left( 1 - \text{sgn}\left(|\braket{\x}{\x_1}|^q - | \braket{\x}{\x_2} |^q \right) \right),\; q=2.
	\end{align}
For simplicity, we omit the parameter $\theta$ and write $\tilde{\x} = \tilde{\x}(\theta)$ when the meaning is clear. The classification for this trivial example requires quantum state fidelity rather than the real component of the inner product as the distance  measure,  verifying  the  advantage of the proposed method. Since the classification relies on the distance between the training and test data in the quantum feature space, we also choose $c$ as to compare the distance between the test datum and training data of each class. The inner products are $\braket{\tilde{\x}}{\x_1} = i \sin {\left( \frac{\theta}{2} + \frac{\pi}{4} \right)}$, and $\braket{\tilde{\x}}{\x_2} = i \cos {\left( \frac{\theta}{2} + \frac{\pi}{4} \right)}$. According to Eq.~(\ref{eq:DesignedKernel}) the expectation value is

    \begin{align}
	\label{eqn:example_classification_statistics}
        \langle \sigma_z^{(a)}\sigma_z^{(l)}\rangle 
		&= w_1|\braket{\tilde{\x}}{\x_1}|^2 - w_2| \braket{\tilde{\x}}{\x_2} |^2\nonumber\\
		&= w_1 \sin^2 {\left( \frac{\theta}{2} + \frac{\pi}{4} \right)} - w_2 \cos^2 {\left( \frac{\theta}{2} + \frac{\pi}{4} \right)}.
    \end{align}
Thus the swap-test classifier outputs $\tilde y$ that coincides with $c(\tilde\x(\theta))$ $\forall\;\theta$. Note that although we have chosen $q=2$ in this example, the swap-test classifier can correctly assign a new label $\tilde y$ $\forall\; q>0$. In contrast, the Hadamard classifier will have the classification expectation value (see Eq.~(\ref{eqn:hadamard_kernel}))
	\begin{equation}
	\label{eqn:example_classification_statistics_hadamard}
	\langle \sigma_z^{(a)}\sigma_z^{(l)}\rangle = w_1 \Re \braket{\tilde{\x}}{\x_1} - w_2 \Re \braket{\tilde{\x}}{\x_2}  = 0.
	\end{equation}
Thus in this example, for any test data parameterized by $\theta$, the Hadamard classifier cannot find the new label $\tilde y$. This data set will be used throughout the paper for demonstrating all subsequent results. Moreover, since the non-uniform weights merely create a systematic shift of the expectation value (see Methods), without loss of generality, we use $w_1 = w_2 = 1/2$ in all examples throughout the manuscript.	Using the above example data set, we illustrate the sharpening of the classification as $n$ increases in Fig.~\ref{fig:classifier_experiment_copies}.	

There are several interesting remarks on the result described by Eq.~(\ref{eq:DesignedKernel}). First, since the cross-terms of the index qubit cancel out, dephasing noise acting on the index qubit does not alter the final result. The same argument also holds for the label qubit. Moreover, the same result can be obtained with the index and label qubits initialized in the classical state as $\sum_m w_m\ket{y_m}\bra{y_m}\otimes \ket{m}\bra{m}$, where $\sum_m  w_m=1$. 
In fact, since the classification is based on measuring the $\sigma_z$ operator on ancilla and label qubits, our algorithm is robust to any error that effectively appears as Pauli error on the final state of them. It is straightforward to see that any Pauli error that commutes with $\sigma_z^{(a)}\sigma_z^{(l)}$ does not affect the measurement outcome. When a Pauli error does not commute with the measurement operator, such as a single-qubit bit flip error on the ancilla or the label qubit, the measurement outcome becomes $(1-2p)\langle \sigma_z^{(a)}\sigma_z^{(l)}\rangle$, where $p$ is the error rate. This result is due to the fact that Pauli operators either commute or anti-commute with each other. This error can be easily circumvented since the classification only depends on the sign of the measurement outcome as shown in Eq.~(\ref{eq:y-tilde}), as long as $p<1/2$. The same level of the classification accuracy as that of the noiseless case can be achieved by repeating the measurement $O(1/(1-2p)^2)$ times. Also, any error that effectively appears at the end of the circuit on any other qubits does not affect the classification result. Second, as the number of copies of training and test data approaches a large number, we find the limit,
	\begin{equation}
	\label{eqn:limit _probability}
	\lim_{n\rightarrow\infty}\langle \sigma_z^{(a)}\sigma_z^{(l)}\rangle
	= \sum_{m}^M (-1)^{y_m}w_m\delta(\tilde{\x} - \x_m).
	\end{equation}
Therefore, as the number of data copies reaches a large number, the classifier assigns a label to the test data approximately by counting the number of training data to which the test data exactly matches.
\subsection*{Kernel construction from a product state}\label{subsec:quantum-state-preparation}
The classifiers discussed thus far require the preparation of a specific initial state structure. Full state preparation algorithms are able to produce the desired state~\cite{QRAMPhysRevLett.100.160501,QRAMPhysRevA.78.052310,PhysRevA.86.010306,ffqram,1998quant.ph..7053V,PhysRevA.64.014303,2002quant.ph..8112G,2004quant.ph..7102K,Mottonen.QIC.2005,PhysRevA.73.012307,PhysRevA.83.032302,2016PhRvA..93c2318I,PhysRevA.97.052329}.
However, all such approaches implicitly assume knowledge of the training and testing data before preparation, and some of the procedures need classical calculation during a pre-processing step. In this section, we present the implementation of the swap-test classifier when training and test data are encoded in different qubits and provided as a product state. In this case, the classifier does not require knowledge of either training and test data. The input can be intrinsically quantum, or can be prepared from the classical data by encoding training and test data on a separate register. The label qubits can be prepared with an $X^{y_m}$ gate applied to $\ket{0}$.

\begin{figure}[ht]
		\centering
		\includegraphics[width=0.98\columnwidth]{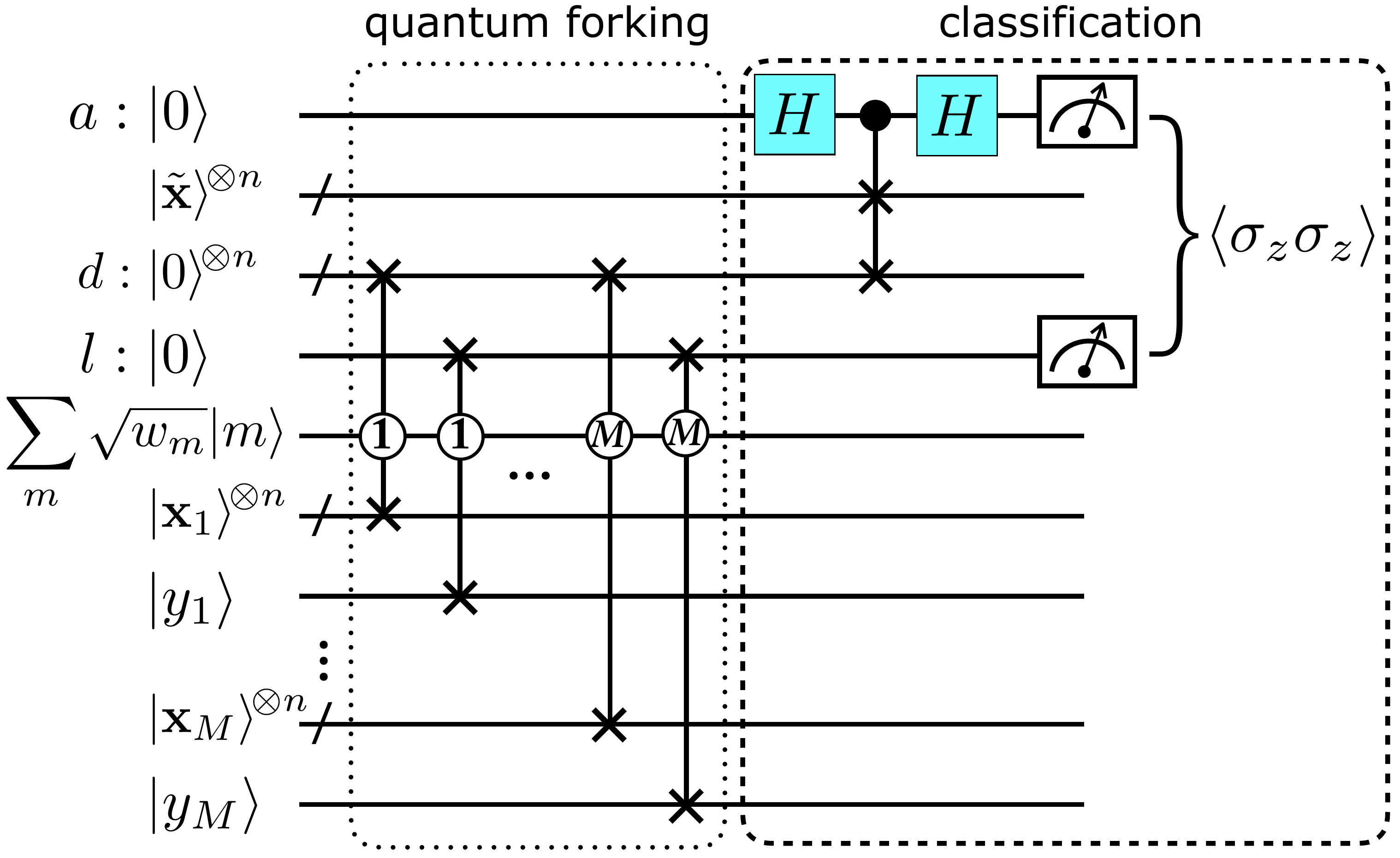}
		\caption{The swap-test classifier with quantum forking for state preparation when the test data, the training data, and the labels are given as a product state.\label{fig:SwapTestClassifier as Product-State}}
	\end{figure}

Given the initial product state, the quantum state required for the swap-test classification can be prepared systematically via a series of controlled-swap gates controlled by the index qubits, which is also provided on a separate register, initially uncorrelated with the reset of the system. The underlying idea is to adapt \textit{quantum forking} introduced in Refs.~\cite{ffqram,qfs} to create an entangled state such that each subspace labeled by a basis state of the index qubits encodes a different training data set. For brevity, we denote the controlled-swap operator by $\text{c-}\texttt{swap}(a,b\vert c)$ to indicate that $a$ and $b$ are swapped if the control is $c$. With this notation, the classification can be expressed with the following equations.
\begin{widetext}
    \begin{align}
	\label{eq:state_prep}
		&\sum_m^M\sqrt{w_m}\ket{0}_a\ket{\tilde\x}^{\otimes n}\ket{0}_d^{\otimes n}\ket{0}_l\ket{m}\ket{\x_1}^{\otimes n}\ket{y_1}\ket{\x_2}^{\otimes n}\ket{y_2}\ldots\ket{\x_M}^{\otimes
		n}\ket{y_M}\nonumber \\
		&\xrightarrow{\prod_m\text{c-}\texttt{swap}(l,y_m\vert m)\cdot\text{c-}\texttt{swap}(d,\x_m\vert m)}\sum_m^M\sqrt{w_m}\ket{0}_a\ket{\tilde\x}^{\otimes n}\ket{\x_m}_d^{\otimes n}\ket{y_m}_l\ket{m}\ket{\text{junk}_m}\nonumber \\
		&\xrightarrow{H_a\cdot\text{c-}\texttt{swap}(d,\tilde\x\vert a)\cdot H_a}\ket{\Phi_f^s}=\frac{1}{2}\sum_m^M\sqrt{w_m}(\ket{0}\ket{\psi_{n+}}+\ket{1}\ket{\psi_{n-}})\ket{y_m}\ket{m}\ket{\text{junk}_m},
    \end{align}
\end{widetext}
where $\ket{\text{junk}_m}$ is some normalized product state. Other than being entangled with the junk state, $|\Phi_f^s\rangle$ in Eq.~(\ref{eq:state_prep}) is the same as $|\Psi_f^s\rangle$ derived in Eq.~(\ref{eq:general_state}). Since $\text{tr}(\ketbra{\text{junk}_m}{\text{junk}_m})=1$, the expectation value of an observable $\sigma_z^{(a)}\sigma_z^{(l)}$ is the same as the result shown in Eq.~(\ref{eq:DesignedKernel}). A quantum circuit for implementing the swap-test classifier with the input data encoded as a product state is depicted in Fig.~\ref{fig:SwapTestClassifier as Product-State}.
	
The entire quantum circuit can be implemented with Toffoli, controlled-\texttt{NOT}, X and Hadamard gates with additional qubits for applying multi-qubit controlled operations. Here we assume that the gate cost is dominated by Toffoli and controlled-\texttt{NOT} gates and focus on counting them using the gate decomposition given in Ref.~\cite{Nielsen:2011:QCQ:1972505}. Note that a Toffoli gate can be further decomposed to one and two qubit gates with six controlled-\texttt{NOT} gates. In total, $n(M+2)\lceil \log_2(N)\rceil+2\lceil \log_2(M)\rceil+M+1$ qubits, $n(M+1)\lceil\log_2(N)\rceil+M\left( 2\lceil\log_2(M)\rceil-1\right)$ Toffoli gates, and $2\left( n(M+1)\lceil\log_2(N)\rceil+M\right)$ controlled-\texttt{NOT} gates are needed. More details on the qubit and gate count can be found in Supplementary Note II. Due to the linear dependence on $M$ and logarithmic dependence on $N$ in the number of gates and qubits, we expect our algorithm to be practically useful for machine learning problems that involve a small number of training data but large feature space. As an example, for $n=1$, the number of qubits, Toffoli and controlled-\texttt{NOT} gates needed for 16 training data with 8 features are 79, 163, and 134. For 16 training data with 16 features, these numbers increase to 97, 180, 168. For 32 training data with 8 features, these numbers become 145, 387 and 262. These numbers suggest that a quantum device with an order of 100 qubits and with an error rate of a Toffoli or a controlled-\texttt{NOT} gate to an arbitrary set of qubits being less than about $10^{-3}$ can implement interesting quantum binary classification tasks. Due to the aforementioned robustness to some errors that effectively appear on the final state, we expect the requirement on the gate fidelity to be relaxed. To our best knowledge, currently available quantum devices does not satisfy the above technical requirement. Nevertheless, with an encouragingly fast pace of improvement in quantum hardware~\cite{Google_QS,ion_trap_BM}, we expect interesting machine learning tasks can be performed using our algorithm in the near future.

\subsection*{The connection to the Helstrom measurement}
The swap-test classifier turns out to be an adaptation of the measurement of a Helstrom operator, which leads to the optimal detection strategy for deciding which of two density operators $\rho_0$ or $\rho_1$ describes a system. The quantum kernel shown in Eq.~(\ref{eq:DesignedKernel}) is equivalent to measuring the expectation value of an observable,
	\begin{equation}
	    \mathcal{A}=\!\!\sum_{m|y_m=0}\!\! w_m\left(\ketbra{\x_m}{\x_m}\right)^{\otimes n}-\!\!\sum_{m|y_m=1}\!\! w_m\left(\ketbra{\x_m}{\x_m}\right)^{\otimes n},
	\end{equation}
on $n$ copies of $\ket{\tilde{\x}}$. This can be easily verified as follows:
	\begin{align}
	    \langle \mathcal{A}\rangle&=\text{tr}\left(\mathcal{A}\ketbra{\tilde{\x}}{\tilde{\x}}^{\otimes n} \right)\nonumber \\
	    &=\text{tr}\left\lbrack\sum_{m}^M(-1)^{y_m}w_m\left(\ketbra{\x_m}{\x_m}\cdot \ketbra{\tilde{\x}}{\tilde{\x}}\right)^{\otimes n}\right\rbrack\nonumber \\
	    &=\sum_{m=1}^M(-1)^{y_m}w_m|\braket{\tilde\x}{\x_m}|^{2n}.
	\end{align}
The above observable can also be written as a Helstrom operator $p_0\rho_0-p_1\rho_1$, where $\rho_i$ represents a hypothesis under a test with the prior probability $p_i$ in the context of quantum state discrimination, by defining $\rho_i=\sum_{m|y_m=i}(w_m/p_i)\ketbra{\x_m}{\x_m}^{\otimes n}$, where $\sum_{m|y_m=i}w_m/p_i=1$ and $p_0+p_1=1$. In this case, measuring the expectation value of $\mathcal{A}$ is equivalent to measuring the expectation value of a Helstrom operator with respect to the test data. The ability to implement the swap-test classifier without knowing the training data via quantum forking leads to a remarkable result that the measurement of a Helstrom operator can also be performed without \textit{a priori} information of target states.
\subsection*{Experimental and Simulation Results}
To demonstrate the proof-of-principle, we applied the swap-test classifier to solve the toy problem of Eq.~(\ref{eqn:data_example}) using the IBM Q 5 Ourense (\texttt{ibmq\_ourense})~\cite{ibm_q_experience} quantum processor. Since $n=1$ in this example, five superconducting qubits are used in the quantum circuit. The number of elementary quantum gates required for realizing the example classification is 27: 14 single-qubit gates and 13 controlled-\texttt{NOT} gates (see Supplementary Fig.~6), which is small enough for currently available noisy-intermediate scale quantum (NISQ) devices. 
    
\begin{figure}[t]
    \centering
	\includegraphics[width=1.0\columnwidth]{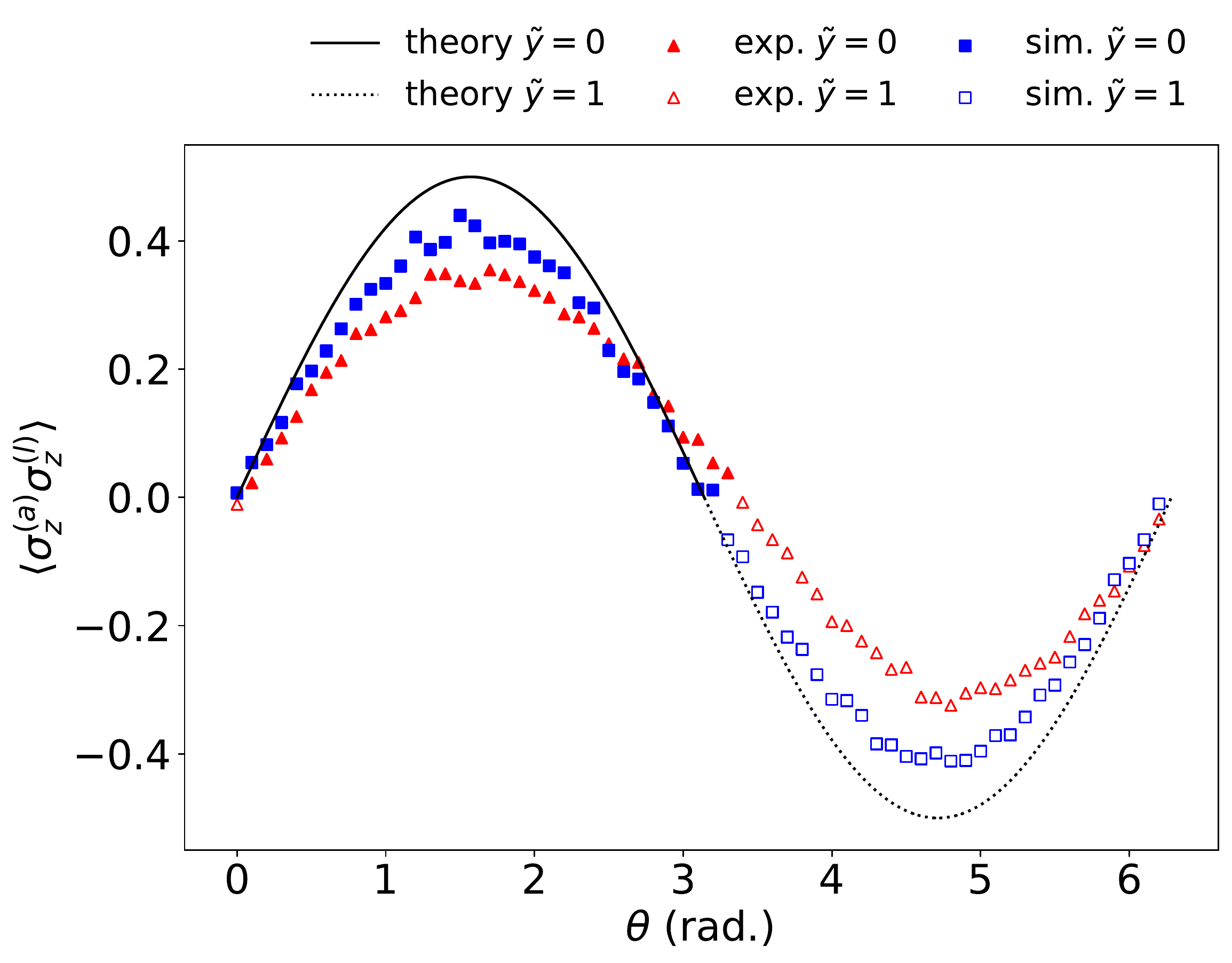}
    \caption{Classification of the toy problem outlined in Eqs.~(\ref{eqn:data_example}) and~(\ref{eqn:example_classification_statistics}) vs. $\theta$. The test data is classified as 0 (1) if the expectation value, $\langle \sigma_z^{(a)} \sigma_z^{(l)} \rangle$, is positive (negative). The experimental result (red triangles) is compared to simulation with a noise model relevant to currently available quantum devices (blue squares) and to the theoretical values (black line). \label{fig:5}}
\end{figure}

The experimental results are presented with triangle symbols, and compared to the theoretical values indicated by solid and dotted lines in Fig.~\ref{fig:5}. 
Albeit having an amplitude reduction of a factor of about $0.65$ and a small phase shift in $\theta$ of about $2\degree$, the experimental result qualitatively agrees well with the theory. 
We performed simulations of the experiment using the IBM quantum information science kit (\texttt{qiskit})~\cite{Qiskit} with realistic device parameters and a noise model in which single- and two-qubit depolarizing noise, thermal relaxation errors, and measurement errors are taken into account. The noise model provided by \texttt{qiskit} is detailed in Supplementary Note III.
The relevant parameters used in simulations are typical data for \texttt{ibmq\_ourense}, and are listed in Supplementary Table I. 
The simulation results are shown as blue squares in Fig.~\ref{fig:5} and we find amplitude reduction of a factor of about $0.82$ with a negligible phase shift.
The difference between simulation and experimental results can be attributed to time-dependent noise, various cross-talk effects~\cite{1908.09855}, and non-Markovian noise.

Despite imperfections, the experiment demonstrates that the swap-test classifier predicts the correct class for most of the input $\tilde{\x}$ (about 97\% of the points sampled in this experiment) in this toy problem. Supplementary Information reports experimental and simulation results obtained from various cloud quantum computers provided by IBM, repeated several times over months. In summary, all results agree qualitatively well with the theory and manifest successful classification with high probabilities.

\section*{Discussion}
We presented a quantum algorithm for constructing a kernelized binary classifier with a quantum circuit as a weighted power sum of the quantum state fidelity of training and test data. The underlying idea of the classifier is to perform a swap-test on a quantum state that encodes data in a specific form. The quantum data subject to classification can be intrinsically quantum or classical information that is transformed to a quantum feature space. We also proposed a two-qubit measurement scheme for the classifier to avoid the classical pre-processing of data, which is necessary for the method proposed in Ref.~\cite{QML_Maria_Francesco}. Since our measurement uses the expectation value of a two-qubit observable for classification, it opens up a possibility to apply error mitigation techniques~\cite{PhysRevLett.119.180509,PhysRevX.8.031027} to improve the accuracy in the presence of noise without relying on quantum error correcting codes. We also showed an implementation of the swap-test classifier with training and test data encoded in separate registers as a product state by using the idea of quantum forking. This approach bypasses the requirement of the specific state preparation and the prior knowledge of data at the cost of increasing the number of qubits linearly with the size of the data. The downside of this approach, which may limit its applicability, is the use of many qubits which must be able to interact with each other. The exponential function of the fidelity approaches to the Dirac delta function as the number of data copies, and hence the exponent, increases to a large number. In this limit, the test data is assigned to a class which contains a greater number of training data that is identical to the test data. An intriguing question that stems from this observation is whether such behaviour of the classifier with respect to the number of copies of quantum information is related to a consequence of the classical limit of quantum mechanics.

Our results are imperative for applications of quantum feature maps such as those discussed in Refs.~\cite{PhysRevLett.122.040504,Havlicek2019}.
In this setting, data will be mapped into the Hilbert space of a quantum system, i.e., $\Phi: \RR^d \rightarrow \mathcal{H}$. Then our classifier can be applied to construct a feature vector kernel as $| \langle \Phi(\x)|\Phi(\x_m) \rangle |^{2n} := K(\x, \x_m)$. Given the broad applicability of kernel methods in machine learning, the swap-test classifier developed in this work paves the way for further developments of quantum machine learning protocols that outperform existing methods. While the Hadamard classifier developed in Ref.~\cite{QML_Maria_Francesco} also has the ability to mimic the classical kernel efficiently, only the real part of quantum states are considered. This may limit the full exploitation of the Hilbert space as the feature space. Furthermore, quantum feature maps are suggested as a candidate for demonstrating the quantum advantage over classical counterparts. It is conjectured that kernels of certain quantum feature maps are hard to estimate up to a polynomial error classically~\cite{Havlicek2019}. If this is true, then the ability to construct a quantum kernel via quantum forking and the swap-test can be a valuable tool for solving classically hard machine learning problems.
	
We also showed that the swap-test classification is equivalent to measuring the expectation value of a Helstrom operator. According to the construction of the swap-test classifier based on quantum forking, this measurement can be performed without knowing the target states under hypothesis in the original state discrimination problem by Helstrom~\cite{Helstrom1969}. The derivation of the measurement of a Helstrom operator from the swap-test classifier motivates future work to find the fundamental connection between the kernel-based quantum supervised machine learning and the well-known Helstrom measurement for quantum state discrimination. Another interesting open problem is whether the Helstrom measurement is also the optimal strategy for classification problems.

During the preparation of this manuscript, we became aware of the independent work by Sergoli et al.~\cite{cagliari.hqc}, in which a quantum-inspired classical binary classifier motivated by the Helstrom measurement was introduced and was verified to solve a number of standard problems with promising accuracy. They also independently found an effect of using copies of the data and reported an improved classification performance by doing so. This again advocates the potential impact of the swap-test classifier with a kernel based on the power summation of quantum state fidelities for machine learning problems.

Other future works include the extension of our results to constructing other types of kernels, the application to quantum support vector machines~\cite{PhysRevLett.113.130503}, and designing a protocol to enhance the classification by utilizing non-uniform weights in the kernel.
\section*{Methods}
\begin{figure}[t]
\centering
\includegraphics[width=1.0\columnwidth]{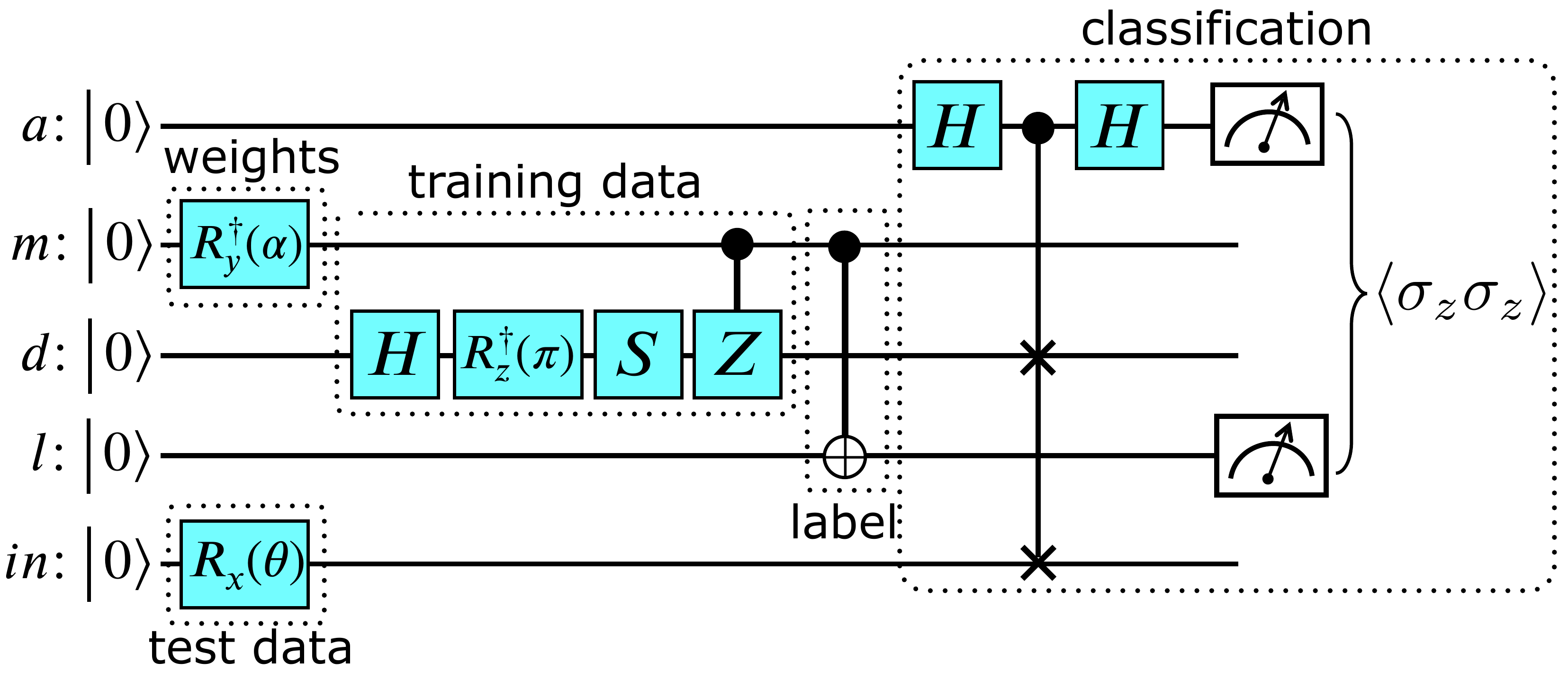}
\caption{\label{fig:NisqExperimentCircuit}The circuit implementing the swap-test classifier on the example data set given in Eq.~(\ref{eqn:data_example}).}
\end{figure}

The quantum circuit implementing the problem of Eq.~(\ref{eqn:data_example}) is shown by Fig.~\ref{fig:NisqExperimentCircuit} where $\alpha$ denotes the angle to prepare the index qubit to accommodate the weights $w_1$ and $w_2$, and $\theta$ is the parameter of the test datum. The experiment applied $\theta$ from 0 to $2\pi$ in increments of 0.1. The experiment for each $\theta$ is executed with 8129 shots to collect measurement statistics. All experiments are performed using a publicly available IBM quantum device consisting of five superconducting qubits, and we used the IBM quantum information science kit (\texttt{qiskit}) framework~\cite{Qiskit} for circuit design and processing.

Superconducting quantum computing devices that are currently available via the cloud service, such as those used in this work, have limited coupling between qubits. The challenge of rewriting the quantum circuit to match device constraints can be easily addressed for a small number of qubits and gates. The quantum circuit layout with physical qubits of the device is shown in Supplementary Information. A minor challenge to be addressed is that each quantum operation of an algorithm must be decomposed into native gates that can be realized with the IBM quantum device. This step is done by the pre-processing library of \verb|qiskit|. The final circuit that is executed on the device consists of 14 single-qubit gates and 13 controlled-\texttt{NOT} gates and is shown in Supplementary Fig.~6. The measurement statistics are gathered by repeating the two-qubit projective measurement in the $\sigma_z$ basis. The expectation value is calculated by $\langle \sigma_z^{(a)} \sigma_z^{(l)} \rangle = \frac{1}{8192}\left(c_{00} - c_{01} - c_{10} + c_{11}\right)$, where $c_{al}$ denotes the count of measurement when the ancilla is $a$ and the label is $l$.

The noise model that we use for classical simulation of the experiment is provided as the \textit{basic model} in \texttt{qiskit} and is explained in detail in Supplementary Information. In brief, the device calibration data and parameters, such as $T_1$ and $T_2$ relaxation times, qubit frequencies, average gate error rate, read-out error rate have been extracted from the API for \texttt{ibmq\_ourense} with the \textit{calibration date} 2019-09-29 11:48:14 UTC. The simulation also requires the gate times, which can be extracted from the device data. As mentioned above, the \textit{basic error model} does not include various cross-talk effects, drift, and non-Markovian noise. Supplementary Information details how the device data and parameters are used in the simulation, and lists the values.

The versions---as defined by PyPi version numbers---we used for this work were 0.7.0 - 0.10.0.

\subsection*{Data availability}
The datasets generated during and/or analysed during the current study are available on the GitHub repository~\cite{SupplementaryRepository}.

\section*{Acknowledgements}
We acknowledge use of IBM Q for this work. The views expressed are those of the authors and do not reflect the official policy or position of IBM or the IBM Q team. This research is supported by the National Research Foundation of Korea (Grant No. 2019R1I1A1A01050161 and 2018K1A3A1A09078001), by the Ministry of Science and ICT, Korea, under an ITRC Program, IITP-2019-2018-0-01402, and by the South African Research Chair Initiative of the Department of Science and Technology and the National Research Foundation. We thank Spiros Kechrimparis for stimulating discussions on the Helstrom measurement. We acknowledge use of the IBM Q for this work. The views expressed are those of the authors and do not reflect the official policy or position of IBM or the IBM Q team.

\section*{Author contributions statement}
C.B. and D.K.P. contributed equally to this work. C.B. and D.K.P designed and analysed the model. C.B. conducted the simulations and the experiments on the IBM Q. All authors reviewed and discussed the analyses and results, and contributed towards writing the manuscript. F.P. is the corresponding author.\\
\linebreak
\textbf{Competing interests} The authors declare no competing interests.



\setcounter{equation}{0}
\setcounter{table}{0}
\setcounter{figure}{0}
\setcounter{section}{0}
\renewcommand{\theequation}{S\arabic{equation}}
\renewcommand{\figurename}{SUPPLEMENTARY FIG.}
\renewcommand{\tablename}{SUPPLEMENTARY TABLE}
\include{circuits}
\onecolumngrid
\bigskip
\begin{center}
{\bf Supplementary Information: Quantum classifier with tailored quantum kernels}
\end{center}

\section{Supplementary Note: Reducing the number of experiments}

The post-measurement scheme of Ref.~\cite{SUPPL_QML_Maria_Francesco} succeeds with the classification if the ancilla is in the ground state (i.e. $\ket{0}$), where the probability to be in the ground ($a=0$) and excited ($a=1$) state is given by $p_a=\sum_{m=1}^M w_m (1 + (-1)^a\text{Re}\braket{\psi_{\x_m}}{\psi_{\tilde{\x}}})/2$. The post-selection scheme will take a toll on the number of experiments that have to be discarded, in particular if $p_0$ is small. This can be circumvented by standardizing the data, i.e., having mean $\mathbf{0}$ and covariance $\mathbf{1}$~\cite{SUPPL_QML_Maria_Francesco}. In this case, $p_0 = p_1$ is attained in the limit $M\rightarrow\infty$ as the number of samples grow. For a proof of this statement, observe that
\begin{align}
    \label{eqn:1}
    |p_0 - p_1| = \frac{1}{2} \left| \sum_{m=1}^M w_m (1+\text{Re}\braket{\x_m}{\tilde{\x}}) - w_m (1-\text{Re}\braket{\x_m}{\tilde{\x}}) \right|
    = \left|\sum_{m=1}^M w_m \text{Re}\braket{\x_m}{\tilde{\x}} \right|.
\end{align}
Let $X, Y \sim \mathcal{N}(0, 1)$ be two independent Gaussian random variables. Then we know that $X \cdot Y \sim c_1 Q - c_2 R$ where $Q, R \sim \chi^2(1)$ and
\begin{align*}
    c_1 &= \frac{Var(X+Y)}{4} = \frac{1}{2}, \\
    c_2 &= \frac{Var(X-Y)}{4} = \frac{1}{2}.
\end{align*}
As $Var(X) = Var(Y)$, both $Q$ and $R$ are independent. The expectation value is given by $\mathbb{E}[XY] = c_1 - c_2 = 0$. Given $\mathbf{X} = (X_1, \ldots, X_d)$ and $\mathbf{Y} = (Y_1, \ldots, Y_d) \sim \mathcal{N}_d(\mathbf{0}, \mathbf{1})$ $d$-dimensional multivariate standard Gaussian random vectors, we know that each of the marginal distributions $X_i$ and $Y_i$ are independent and identically distributed (i.i.d.) and standard uni-variate Gaussian random variables. Therefore $\mathbbm{E}\left[ \mathbf{X} \cdot \mathbf{Y} \right] = \mathbb{E}[X_1 Y_1 + \cdots + X_d Y_d] = 0$. Now if $\tilde{\x}$ is a realization of $\mathbf{X}$ and $(\x_m)$ are $M$ realizations of $\mathbf{Y}$, then $(\braket{\tilde{\x}}{\x_m})$ are $M$ realizations of $\mathbf{X}\cdot\mathbf{Y}$. In Ref.~\cite{SUPPL_QML_Maria_Francesco} it was assumed that $w_m = 1/M$, and therefore we find that $p_0 - p_1$ is indeed the mean of the series of inner products. This shows that $|p_0 - p_1| \rightarrow 0$ as $M\rightarrow\infty$. Now, even if $p_0$ is very small given raw data, once pre-processed, this allows for the post-selection to succeed with the probability close to 1/2. Nevertheless, since $p_1$ is also close to 1/2, half of the experiments are discarded in the classification.
As a consequence, the two-qubit measurement introduced in the main manuscript will result in reducing the number of experiments by about a factor of $2$ if the data is real-valued and approximately multivariate normal.

As briefly discussed in the main text, the same argument does not apply to the \textit{swap-test} classifier, as we have
\begin{align}
    \label{eqn:2}
    |p_0 - p_1| = \sum_{m=1}^M w_m |\braket{\x_m}{\tilde{\x}}|^{2n}
\end{align}
and the expectation value, for standardized data, will always be positive. Indeed, one must argue that $p_0$ will in expectation be greater than $p_1$. Hence the expectation value measurement does not provide the factor of two speed-up with respect to the number of experiments. Nevertheless, all experiments contributes to the classification. As such, we conclude that the two-qubit expectation value measurement is an improvement from the post-selection scheme in both cases, the \textit{Hadamard} and the \textit{swap-test} classifier.

\section{Supplementary Note: Qubit and gate counts for implementing the swap-test classifier with a product input state and quantum forking}
Using the gate decomposition given in Ref.~\cite{SUPPL_Nielsen:2011:QCQ:1972505}, the number of qubits and gates needed for implementing the quantum circuit of the swap-test classifier for which the input state is given as a product state (Fig.~4 of the main manuscript) can be counted as follows. Recall that the number of the training data is $M$, the dimension of the features is $N$ and that the number of identical copies being used is $n$. A swap operation used in quantum forking controlled by $\lceil \log_2(M) \rceil$ index qubits requires $\lceil \log_2(M) \rceil-1$ ancilla qubits. The circuit also requires one ancilla qubit, $2n\lceil \log_2(N) \rceil$ qubits for the test data and the space to register the training data, one qubit for registering labels, $Mn\lceil \log_2(N) \rceil$ training data qubits, and $M$ label qubits. In total, $n(M+2)\lceil \log_2(N)\rceil+2\lceil \log_2(M)\rceil+M+1$ qubits are needed. The entire scheme can be implemented with Toffoli, controlled-\texttt{NOT}, X and Hadamard gates. We assume that the gate cost is dominated by Toffoli and controlled-\texttt{NOT} gates and focus on counting them. Note that a Toffoli gate can be further decomposed to one and two qubit gates with six controlled-\texttt{NOT} gates.

Quantum forking requires $M$ gates that swap $n\lceil \log_2(N) \rceil$ qubits with the empty data register qubits (labeled as $d$ in Fig.~4 of the main manuscript), and $M$ gates that swap label qubits with the empty label register qubit (labeled as $l$ in Fig.~4 of the main manuscript), controlled by $\lceil \log_2(M) \rceil$ index qubits. A multi-qubit controlled-swap operation can be implemented by using $2(\lceil \log_2(M) \rceil-1)$ extra Toffoli gates, X gates for selecting a control bit string of the index qubits, and a swap gate controlled by one qubit. Since each single-qubit controlled-swap operation can be decomposed as a Toffoli and two controlled-\texttt{NOT} gates (see Supplementary Fig.~\ref{fig:cswap2cnotToff}), the total number of Toffoli and controlled-\texttt{NOT} gates required for quantum forking is $M\left( 2\lceil\log_2(M)\rceil+n\lceil\log_2(N)\rceil-1\right)$ and $2M\left( n\lceil\log_2(N)\rceil+1\right)$, respectively. The classification step requires $n\lceil \log_2(N) \rceil$ Toffoli gates, $2n\lceil \log_2(N) \rceil$ controlled-\texttt{NOT} gates, and two Hadamard gates. In summary, $n(M+1)\lceil\log_2(N)\rceil+M\left( 2\lceil\log_2(M)\rceil-1\right)$ Toffoli gates and $2\left( n(M+1)\lceil\log_2(N)\rceil+M\right)$ controlled-\texttt{NOT} gates are needed in total.

\begin{figure}[h]
	\centering
	\includegraphics[width=0.3\textwidth]{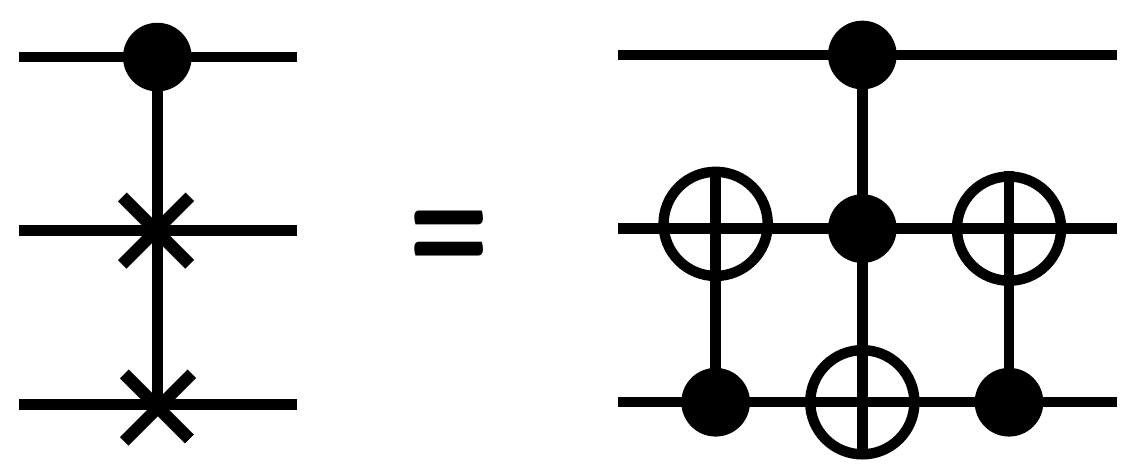}
	\caption{Decomposition of a controlled-swap gate into controlled-\texttt{NOT} and Toffoli gates. \label{fig:cswap2cnotToff}}
\end{figure}
\section{Supplementary Note: Details on simulation and experiment with IBM Quantum Experience}

\subsection{Preliminaries}

This section describes details of simulations and experiments presented in the main manuscript and references to the data. For all simulations and experiments, we used IBM quantum information science kit (\texttt{qiskit}) framework~\cite{SUPPL_Qiskit}. The versions---as defined by PyPi version numbers---we used were 0.7.0 - 0.10.0.

As grounds of our technical endeavors we use the classification example Eq. (11) of the main manuscript:
\begin{equation}
	\ket{\x_1} = \frac{i}{\sqrt{2}} \ket{0} + \frac{1}{\sqrt{2}} \ket{1},\; y_1 = 0,\;
	\ket{\x_2} = \frac{i}{\sqrt{2}} \ket{0} - \frac{1}{\sqrt{2}} \ket{1},\; y_2 = 1,\;
	\ket{\tilde{\x}(\theta)} = \cos{\frac{\theta}{2}} \ket{0} - i \sin{\frac{\theta}{2}} \ket{1}.
\end{equation}
In case of a binary classification problem, a true label function $c$ must be given which assigns each data sample $\x$ a label $0,1$. For learning algorithms that use a similarity measure as basis this is simply given by
\begin{align*}
    c(\x) = \begin{cases}
    0,\qquad w_1 D(\x,\x_1) > w_2 D(\x,\x_2) \\
    1,\qquad w_1 D(\x,\x_1) < w_2 D(\x,\x_2) \\
    \frac{1}{2}, \qquad w_1 D(\x,\x_1) = w_2 D(\x,\x_2)
    \end{cases}
\end{align*}
or equivalently,
\begin{equation}
    c(\x) = \frac{1}{2} \left( 1 - \text{sgn}\left( w_1 D(\x,\x_1) - w_2 D(\x,\x_2) \right) \right).
\end{equation}
The classification is thus dependent on a similarity measure. In general a similarity measure, as given above, is a real-valued function $D$. In a quantum setting, the common similarity measures are the quantum state fidelity or the state overlap, i.e., the inner product $D(\cdot, \cdot) = \braket{\cdot}{\cdot}$. A very common similarity measure reminiscent to the state overlap is the cosine similarity $D(\x_1, \x_2) = \x_1 \cdot \x_2/(\|\x_1\| \|\x_2\|)$ for real valued $d$ dimensional vectors. It is quite interesting to note that the \textit{Hadamard} classifier favors the latter while the \textit{swap-test} protocol favors the former. The main focus of our work relates to quantum feature maps projecting real-valued data into a high dimensional (quantum) Hilbert space, and therefore invoking the need for a natural similarity measure such as the quantum state fidelity. Therefore the true label function $c$ is defined in our case with the state fidelity as similarity measure.

Applying the classifiers to the above problem, the expectation values of the two-qubit observable used for the \textit{swap-test} and the \textit{Hadamard} classifier are
\begin{equation}
    \langle \sigma_z^{(a)}\sigma_z^{(l)}\rangle 
	= w_1|\braket{\tilde{\x}}{\x_1}|^2 - w_2| \braket{\tilde{\x}}{\x_2} |^2
	= w_1 \sin^2 {\left( \frac{\theta}{2} + \frac{\pi}{4} \right)} - w_2 \cos^2 {\left( \frac{\theta}{2} + \frac{\pi}{4} \right)},
\end{equation}
and
\begin{equation}
    \langle \sigma_z^{(a)}\sigma_z^{(l)}\rangle = w_1 \Re \braket{\tilde{\x}}{\x_1} - w_2 \Re \braket{\tilde{\x}}{\x_2}  = 0,
\end{equation}
respectively (see Eq. (12) and Eq. (13) in the main manuscript).
As $w_1 + w_2 = 1$, we get
$
\langle \sigma_z^{(a)}\sigma_z^{(l)}\rangle = \sin^2 {\left( \frac{\theta}{2} + \frac{\pi}{4} \right)} - w_2
$,
which is a simple translation. For simplicity, we used $w_1 = w_2 = \frac{1}{2}$ in all simulations and experiments. Having the previous discussion in mind we see that the \textit{Hadamard} classifier, favoring the cosine similarity and forcing real-values, will evaluate the test datum equally similar to each of the training samples. This example is of course chosen with the intention to demonstrate that only a classifier with a similarity measure that also takes imaginary values into account, can be useful for future applications of quantum feature maps to the full extent.

As the toy problem defined here includes a parameter $\theta \in (0, 2\pi)$, we need to systematically apply this range of values in the experiment. An equidistant discretization of the interval is done in steps of $0.1$. For each $\theta$, one circuit is transpiled and sent together in one batch (called \texttt{Qobj} in \texttt{qiskit}) to either the simulator or the API (hence the device). Experiments and simulations are executed with 8129 shots to collect measurement statistics.

\subsection{Circuit Design}

The state to prepare is $\ket{\Psi_i^s} = \sum_{m=1}^M\sqrt{w_m}\ket{0}\ket{\tilde{\x}}^{\otimes n}\ket{\x_m}^{\otimes n}\ket{y_m}\ket{m}$. In fact, the toy problem of Eq.~(11) of the main text was chosen to maximize the improvement of classification with respect to the \textit{Hadamard} classifier and to be preparable with only one entangling operation difference. The resulting circuit is depicted in Supplementary Fig.~\ref{supl_fig:NisqExperimentCircuit} for the \textit{swap-test} classifier. We use \texttt{qiskit} to program the circuit using Python (see the code listing in Supplementary Note~\ref{fig:circuit_python_code}). The generic circuit shown in Fig.~4 of the main manuscript is also applied to the example problem defined in Eq.~(11) of the main manuscript.
\begin{figure}[h]
	\centering
	\includegraphics[width=0.6\textwidth]{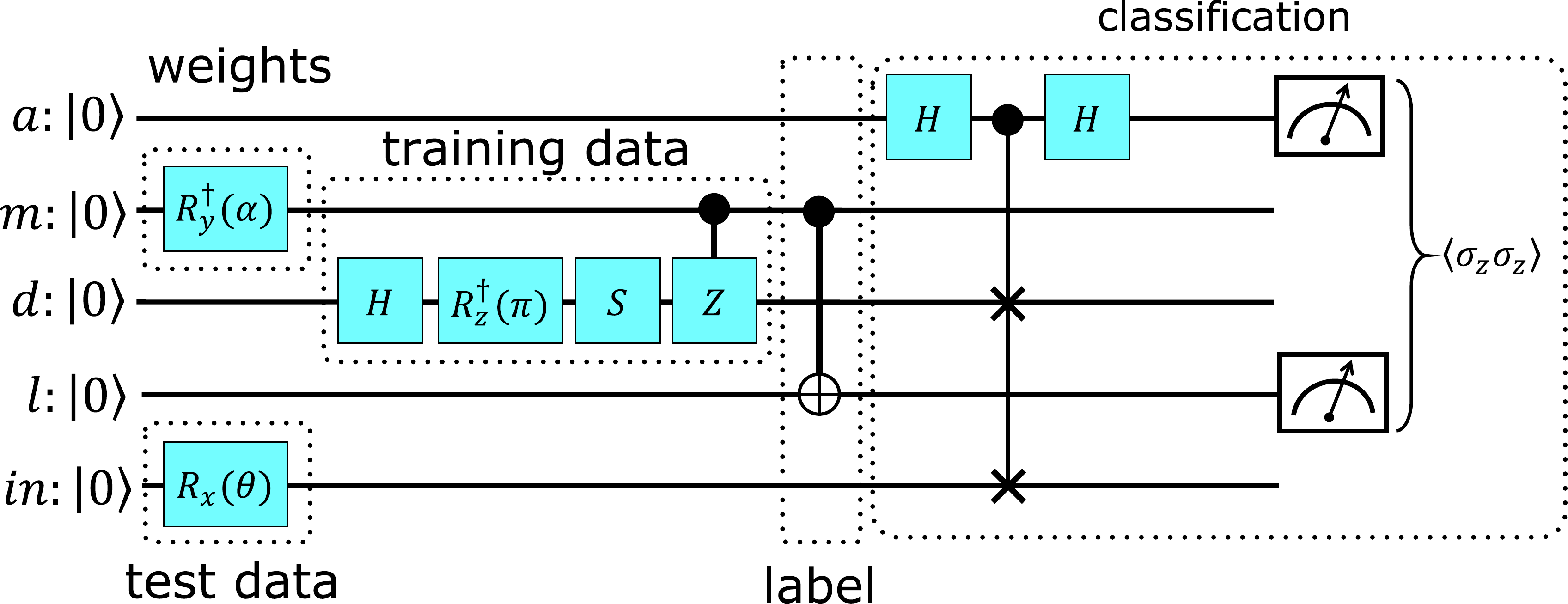}
	\caption{The circuit implementing the \textit{swap-test} classifier on the example. \label{supl_fig:NisqExperimentCircuit}}
\end{figure}

The non-uniform weights $w_1$ and $w_2$ with $w_1 + w_2 = 1$ can be realized by applying a $Y$-rotation on the index register $\ket{m}$ with an angle $\alpha = 2 \sin^{-1}{\left(\sqrt{w_2}\right)}$. The full $n$-copy circuit code is given in the GitHub repository~\cite{SUPPL_SupplementaryRepository}, while an example circuit for $n=10$ is shown in Supplementary Fig.~\ref{fig:TenCopiesHighLevelCircuit}.

Superconducting qubit devices, such as those provided via the IBM cloud, are limited in the coupling between physical qubits. Qubit couplings are needed in order to be able to construct arbitrary unitaries, a natural prerequisite to most useful applications. For a small number of qubits and gates, as in our toy example, we were able to hand-pick a logical-to-physical qubit mapping. An analysis of the requirements shows that there are two groups of coupled logical qubits. The first is ancilla--data--input ($a, d, in$) and the second is index--data--label ($m, d, l$). Given the coupling map of the \texttt{ibmq\_ourense} (see Supplementary Fig.~\ref{fig:coupling_ourense}), we see that the mapping $a \rightarrow q_0,\ m \rightarrow q_3,\ d \rightarrow q_1,\ l \rightarrow q_4,\ in \rightarrow q_2$ allows for the initial data to be encoded into a feasible circuit. In order to apply a controlled-swap operation, we use a decomposition by a controlled-\texttt{NOT}, a Toffoli and a controlled-\texttt{NOT} gate. The Toffoli gate will need a swap gate in order to be able to entangle the train and test data with the ancilla register. For more information see the GitHub repository referenced by~\cite{SUPPL_SupplementaryRepository}, which also includes implementations to various backends provided by IBM as well as a circuit for the Hadamard classifier.
\begin{figure}[h]
	\centering
	\includegraphics[width=0.2\textwidth]{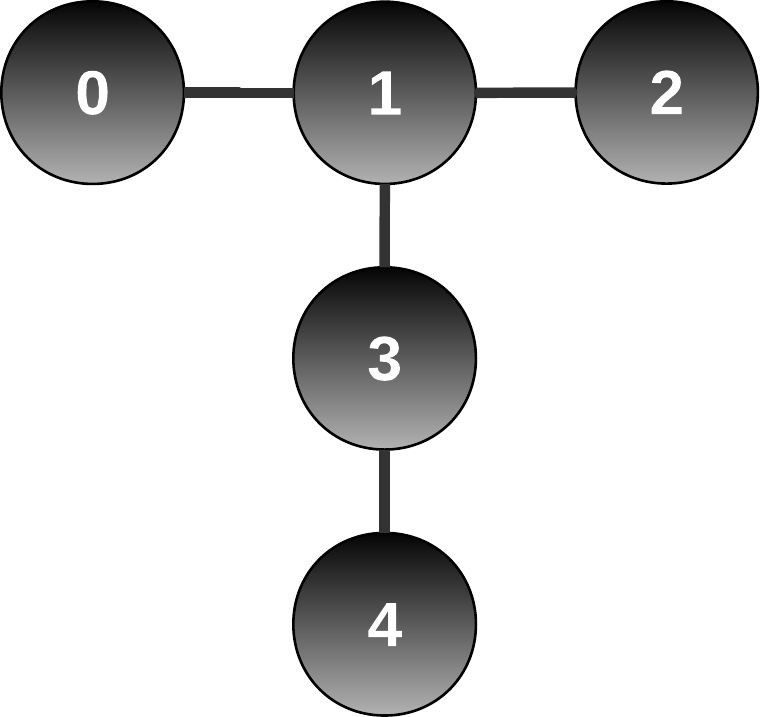}
	\caption{Coupling map of the IBM quantum devices, \texttt{ibmq\_ourense}. Source Ref.~\cite{SUPPL_ibm_q}.} \label{fig:coupling_ourense}
\end{figure}

Each applied quantum operation of an algorithm must be decomposed into native gates that can be realized with the IBM quantum device. An arbitrary single qubit unitary operation can be expressed as
\begin{equation}
U(\theta,\phi,\lambda) = \begin{pmatrix} \cos(\theta/2) & -e^{i\lambda}\sin(\theta/2) \\ e^{i\phi}\sin(\theta/2)
& e^{i\lambda+i\phi}\cos(\theta/2) \end{pmatrix}.
\end{equation}
The native single qubit gates are then given as \verb|u1|$=U(0,0,\lambda)$, \verb|u2|$=U(\pi/2,\phi,\lambda)$, and \verb|u3|$=U(\theta,\phi,\lambda)$. The native two qubit gate is the controlled-\texttt{NOT} (\verb|cx|) operation. The transpilation resolving most of arbitrary unitary operations to the native gates is done by \verb|qiskit| pre-processing involving a so-called \verb|PassManager| that can be configured as needed. As the logical--to--physical qubit mapping was hand-picked the transpilation consists of three passes without a nearest-neighbor constraint resolving pass:
\begin{itemize}
	\item Decompose all non-native gates (\verb|qiskit.transpiler.passes.Unroller|).
	\item Direct \verb|cx| gates according to coupling map (using \verb|qiskit.transpiler.passes.CXDirection|).
	\item Optimize single qubit gates (using \verb|qiskit.transpiler.passes.Optimize1qGates|).
\end{itemize}
Each original quantum circuit is now transformed to a circuit with the qubit arrangement and gate decomposition that are suitable for the experimental constraints. The described procedure is by no means optimal\footnote{Optimality must first be defined and must take into account environmental noise as well as pulse, readout and cross-talk errors. Such calibration data is partially provided but its effects must be modelled first. A fully automated and almost optimal procedure will therefore be a research area of its own, and we will not dive into it at this point.}.
In order to resolve nearest-neighbor constraint one usually applies two-qubit swap operations which can be decomposed into three \texttt{cx} gates. Since the use of three \texttt{cx} gates is usually an expensive operation we tried to minimize the number of swap gates for connecting physically uncoupled qubits logically. The final number of gates after transpiling the \textit{swap-test} classifier (as implemented by the circuit in Supplementary Fig.~\ref{fig:NisqExperimentCircuit}) is at 27 for all values of $\theta$. The fully transpiled quantum circuit is shown in Supplementary Fig.~\ref{fig:transpiled_swaptest_circuit}.
\subsubsection*{Measurement}

In the main manuscript we introduced the measurement of a two-qubit observable, $\sigma_z^{(a)} \sigma_z^{(l)}$, which has two eigenvalues $+1$ and $-1$. We identify the readouts $00$ and $11$ with the eigenvalue $+1$ and $01$ and $10$ with $-1$. Each single experiment thus has two outcomes $+1$ to $-1$, giving rise to a classification estimator $\hat{c}(\tilde{\x}) = 0$ or $\hat{c}(\tilde{\x}) = 1$, respectively. In fact the expectation value of this estimator is
$$
    \mathbbm{E} [\hat{c}(\tilde{\x})] = \frac{1}{N}\left(c_{00} + c_{11} - (c_{01} + c_{10}) \right)
$$
where $c_{al}$ denotes the count of measurement when the ancilla was $a$ and the label was $l$. By construction this is equal to the expectation value of the two-qubit observable, hence $\mathbbm{E} [\hat{c}(\tilde{\x})] = \langle \sigma_z^{(a)} \sigma_z^{(l)} \rangle$, so the choice of $\hat{c}$ is the naturally arising unbiased estimator of the classification. The code listing in Supplementary Note~\ref{fig:expectation_value_python_code} shows how the readout is converted to an estimation of the classification given the number of shots.

\subsection{Simulation with a realistic noise model}\label{subsec:simulation_noise}

The reason to use a simulator with realistic noise lies in the ability to get a close understanding of the experimental results. Therefore it was desired to apply a reasonably relevant but still easy-to-use noise model. The provided \textit{basic error model} of \verb|qiskit| seemed to fit into those requirements. For this reason it was necessary to fully understand the provided noise model in order to understand the results. As such we did an in-depth code analysis of the applied simulator.

Simulations in this work were executed by using \verb|qiskit-aer|, an open source simulator provided by IBM~\cite{SUPPL_Qiskit}, with the noise model option enabled. The basic noise model that is provided with \verb|qiskit-aer| is found in \verb|qiskit.providers.aer.noise.device.models.basic_device_noise_model|. The noise simulation takes device parameters, calibration data, gate time and temperature as input. There are several groups of device information: \textit{device parameters} (frequency of each qubit $f$ in GHz and temperature of device $T$ in K), \textit{device calibration} (average single-qubit gate infidelities $\epsilon$, \verb|cx| gate error rate $\epsilon_\texttt{cx}$, the readout error rate $\epsilon_r$ and $T_1, T_2$ relaxation times in $\mu$s) and finally \textit{device gate times} in ns denoted by $T_g(\cdot)$, where the argument is a gate, e.g. \texttt{id}, \texttt{u1}, \texttt{u2}, \texttt{u3}, \texttt{cx}. The error model consists of the following local error channels: readout error, depolarizing error and thermal relaxation error.

The model is briefly summarized with examples in the documentation~\cite{SUPPL_qiskit_documentation}. The noise model is a simplified approximation of the real dynamics of a device, and therefore caution of the applicability is given as the study of noisy quantum devices is an active field of research.
The following analysis was done by code-review of the version 0.1.1 of \verb|qiskt-aer|.

\subsubsection*{Readout Error}
The readout error probability is defined as $p_{jm} = \mathbbm{P}(j|m)$, where $m$ is the actual state and $j$ is the measured outcome ($m,j \in \{0,1\}$), and is denoted by $\epsilon_r \hat{=} \verb|readout_error|$ where $\epsilon_r = p_{jm}$ if $j \not= m$.

\subsubsection*{Depolarizing Error}

The depolarizing channel in the absence of $T_1$ and $T_2$ relaxations (i.e., $T_1=T_2=\infty$) is given by the following. Say the average gate error is given by $\epsilon = 1 - F$ where $F$ is the average gate fidelity. The $n$-dimensional depolarizing channel can be represented by the operator
\[
\mathcal{E}_{dep} = (1-p) I + p D
\]
where $I$ is the identity and $D$ is the completely depolarizing channel. The average gate fidelity is then given by
\[
    F(\mathcal{E}_{dep})    = (1-p) F(I) + p F(D) = (1-p) + \frac{p}{n},
\]
where $F(I) = 1$ and $F(D) = n^{-1} = 1 - p \frac{n - 1}{n}$. Therefore it is true that
\[
    p   = \frac{1-F(\mathcal{E}_{dep})}{\frac{n - 1}{n}} = n \frac{1 - F(\mathcal{E}_{dep})}{n - 1} = n \frac{\epsilon}{n-1},
\]
where $n=2^N$, $N$ is the number of qubits, and $\epsilon \hat{=} \verb|error_param|$.

Next, we scrutinize the case when thermal relaxations are present. Starting with the one-qubit case ($n=2$), given a non-negative gate time denoted by $T_g$ and some non-negative values of $T_1$ and $T_2$ that satisfy $T_2\leq 2T_1$, and with $d = \exp{-T_g/T_1} + 2 \exp{-T_g/T_2}$ then
\[
    p = 1 + 3 \frac{2 \epsilon - 1}{d}.
\]
For the two-qubit depolarizing probability ($n=4$), given some non-negative values of $T_{i1}$ and $T_{i2}$ that satisfy $T_{i2}\leq 2 T_{i1}$, where $i\in\lbrace 0,1\rbrace$ labels the qubit, and with $\tau_{ik} = \exp{-\frac{T_g}{T_{ik}}}$ ($k=1,2$),
\[
    d = \tau_{01} + \tau_{11} + \tau_{01} \tau_{11} + 4 \tau_{02} \tau_{12} + 2 (\tau_{02} + \tau_{12}) + 2 (\tau_{11} \tau_{02} + \tau_{01} \tau_{12}),
\]
where $T_g$ is the gate time. Then the depolarizing probability is
\[
    p_2 = 1 + 5\frac{4\epsilon - 3}{d}.
\]
Kraus representation of the depolarizing channel is given by the Kraus operators
\[
\mathcal{E}_n=\lbrace \sqrt{1 - (4^N - 1)p/4^N} I^{\otimes N},\sqrt{p/4^N} \mathcal{P}_j\rbrace
\]
where $\mathcal{P}_j\in \lbrace I,X,Y,Z\rbrace^{\otimes N}\setminus I^{\otimes N}$ denotes an element in the set of $N$-qubit Pauli operators minus the identity matrix.

\subsubsection*{Thermal Relaxation Error}
Thermal relaxation is governed by the relaxation times $T_1, T_2$ with the above constraints and the gate time $T_g$. There is a chance that a \texttt{reset} error (unwanted projection or unobserved measurement) happens, the weight to which state this happens (either towards $\ket{0}$ or $\ket{1}$) is dependent on a value called the \textit{excited state population}, $0 \leq p_e \leq 1$, which is defined as
\[
    p_e = \left( 1 + \exp{\frac{2h f}{ k_B T }} \right)^{-1},
\]
where $T$ is the given temperature in K, $f$ is the qubit's frequency in Hz, $k_B$ is Boltzmann 's constant (eV/K) and $h$ is Planck 's constant (eVs). For the limiting cases we have $p_e = 0$ if the frequency $f \rightarrow \infty$ or temperature $T \rightarrow 0$. The $T_1$ and $T_2$ relaxation error rates can be defined as $\epsilon_{T_1} = \exp{-T_g / T_1}$ and $\epsilon_{T_2} = \exp{-T_g / T_2}$, respectively. From this the defined $T_1$ reset probability is given by $p_\text{reset} = 1 - \epsilon_{T_1}$. Depending on the regime of $T_1$ and $T_2$ there are two different models. If $T_2 \leq T_1$, \texttt{qiskit} implements the thermal relaxation as a probabilistic mixture of \texttt{qobj circuits} from the circuits that implement $I$, $Z$, $\texttt{reset}$ to $\ket{0}$, and $\texttt{reset}$ to $\ket{1}$ with the probabilities
\begin{align*}
    &p_\text{id} = 1 - p_z - p_{r0} - p_{r1},\\
    &p_z = (1 - p_\text{reset}) \left(1 - \epsilon_{T_2}\epsilon^{-1}_{T_1}\right)/2,\\
    &p_{r_0} = (1 - p_e) p_\text{reset},\\
    &p_{r_1} = p_e p_\text{reset},
\end{align*}
respectively. Note that in this case \texttt{qiskit} does not use the Kraus representation. However, the Kraus operators for a \texttt{reset} circuit that projects a given quantum state to $\ket{i}$ can be expressed as
\[
    \mathcal{E}_{r_i}=\lbrace \ket{i}\bra{0} , \ket{i}\bra{1} \rbrace.
\]
If $T_2 > T_1$, then the error channel can be described by a Choi-matrix representation~\cite{SUPPL_1111.6950}. For a quantum channel $\mathcal{E}$, the Choi matrix $\Lambda$ is defined by
\[
    \Lambda =  \sum_{i,j} \ket{i}\bra{j} \otimes \mathcal{E}(\ket{i}\bra{j}).
\]
The evolution of a density matrix with respect to the Choi-matrix is then defined by
\[
    \mathcal{E}(\rho) = \text{tr}_{1}[\Lambda(\rho^T\otimes I)]
\]
where $\text{tr}_1$ is the trace over the first (main) system in which $\rho$ exists. In this thermal relaxation case the Choi-matrix is given by
\[
    \Lambda = \begin{pmatrix}
        1 - p_e p_\text{reset} & 0 & 0 & \epsilon_{T_2} \\
        0 & p_e p_\text{reset} & 0 & 0 \\
        0 & 0 & (1 - p_e) p_\text{reset} & 0 \\
        \epsilon_{T_2} & 0 & 0 & 1 - (1 - p_e) p_\text{reset}
    \end{pmatrix}.
\]
For usability \verb|qiskit-aer| transforms this representation to Kraus maps. If the Choi matrix is Hermitian with non-negative eigenvalues, the Kraus maps are given by $K_\lambda = \sqrt{\lambda} \Phi(v_\lambda)$ where $\lambda$ is an eigenvalue and $v_\lambda$ its eigenvector. Furthermore $\Phi$ is a isomorphism from $\mathbb{C}^{n^2}$ to $\mathbb{C}^{n \times n}$ with column-major order mapping, i.e. $\Phi(\x)_{i,j} = (x_{i + n (j - 1)})$ with $i,j=1, \ldots, n$ and $\x \in \mathbb{C}^{n^2}$. If the Choi matrix has negative eigenvalues or is not Hermitian, a singular value decomposition is applied which leads to two sets of Kraus map. Let $\Lambda = U \Sigma V^\dagger$ be the singular value decomposition with $\Sigma = \operatorname{diag}(\sigma_1, \ldots, \sigma_{n})$ with $\sigma_i \geq 0$. Given $U =(\mathbf{u}_1 | \cdots | \mathbf{u}_n)$, also called the left singular vectors, and $V =(\mathbf{v}_1 | \cdots | \mathbf{v}_n)$, the right singular vectors, then the Kraus maps are computed to be $K_i^{(l)} = \sqrt{\sigma_i} \Phi(\mathbf{u}_i)$ and $K_i^{(r)} = \sqrt{\sigma_i} \Phi(\mathbf{v}_i)$. If left and right Kraus maps aren't equal to each other, i.e. $\mathbf{u}_i \not= \mathbf{v}_i$ for some $i=\{1, \ldots, n\}$, they do not represent a completely positive trace preserving (CPTP) map.

\subsubsection*{Combining Errors and Application to Simulations}

Both error representations, i.e. Kraus maps, are computed independently and then combined by composition.
According to \verb|qiskit-aer| documentation~\cite{SUPPL_qiskit_documentation} the probability of the depolarizing error is set such that the combined gate infidelity of depolarizing and thermal relaxation error is equal to the reported device's average gate infidelity. This is the anchor point of the noise model and the actual measured values of the device.

\subsection{Experimental Results}

The described noise model is applied under \verb|qiskit| as the \textit{basic model} when provided with all of the device's data: qubit frequency, $T_1, T_2$ times, gate and readout error parameters and gate times. If gate times are missing, only depolarizing noise is activated as a gate time of $T_g=0$ results in an equivalent situation as if $T_1=T_2=\infty$. The experiment is prepared as a \texttt{qiskit.qobj.Qobj} which has all 63 circuits (one for each $\theta$ from 0 to $2\pi$ in increments of 0.1) and sent to IBM Q's API to be scheduled for execution.
At the time of the experiment the device parameters from the device's calibration are extracted and saved as part of a Python file with all relevant data. An example of the device data of the experiment on the 2019-09-29 11:48:14 UTC (the file listed in Supplementary Note~\ref{sec:data}) is shown in Supplementary Table~\ref{table:device_data}. Immediately after the experiment this data is applied as the device data to the described noise model. The results of this experiment and simulation are shown in Supplementary Fig.~\ref{fig:4}.
\begin{table}[h]
    \centering
    \subfloat[]{
        \begin{tabular}{ccccccccc}
            Name    & $T_1$ [$\mu$s]& $T_2$ [$\mu$s]& $f$ [GHz] & $\epsilon_r$  & $\epsilon$    & $T_g(u2)$ [ns] & $T$ [K] \\
            \hline
            $Q_0$	&	94.9785 	&	93.2334 	&	4.8195	&	0.0180      &   0.000318 	&	36           & 0.0 \\
            $Q_1$	&	101.3888 	&	36.7808 	&	4.8911	&	0.0180      &   0.000376 	&	36           & 0.0 \\
            $Q_2$	&	179.9652 	&	134.4074 	&	4.7169	&	0.0110      &   0.000290 	&	36           & 0.0 \\
            $Q_3$	&	128.7045 	&	112.0798 	&	4.7890	&	0.0280      &   0.000305 	&	36           & 0.0 \\
            $Q_4$	&	73.0632 	&	39.0842 	&	5.0237	&	0.0310      &   0.000337 	&	36           & 0.0 \\
            \hline
        \end{tabular}
        }\hspace{0.8cm}
    \subfloat[]{
        \begin{tabular}{ccc}
        Name & $\epsilon_{\texttt{cx}}$ & $T_g(cx)$ [ns] \\
        \hline
        $cx_{01}, cx_{10}$	&	0.005685 	&	235	\\
        $cx_{12}, cx_{21}$	&	0.007304 	&	391	\\
        $cx_{13}, cx_{31}$	&	0.011624 	&	576	\\
        $cx_{34}, cx_{43}$	&	0.006492 	&	270	\\
        \hline
    \end{tabular}
}
    \caption{Device data of \texttt{ibmq\_ourense} from the calibration on 2019-09-29 11:48:14 UTC used for the noise model for the simulation matching the experiment shown in Supplementary Fig.~\ref{fig:4}. (a) Single qubit data with an error population temperature of $T$ as well as \texttt{u2} gate times. (b) \texttt{cx} gate times and the average gate infidelity.}
    \label{table:device_data}
\end{table}

\begin{figure}[h]
	\centering
    \includegraphics[scale=0.5]{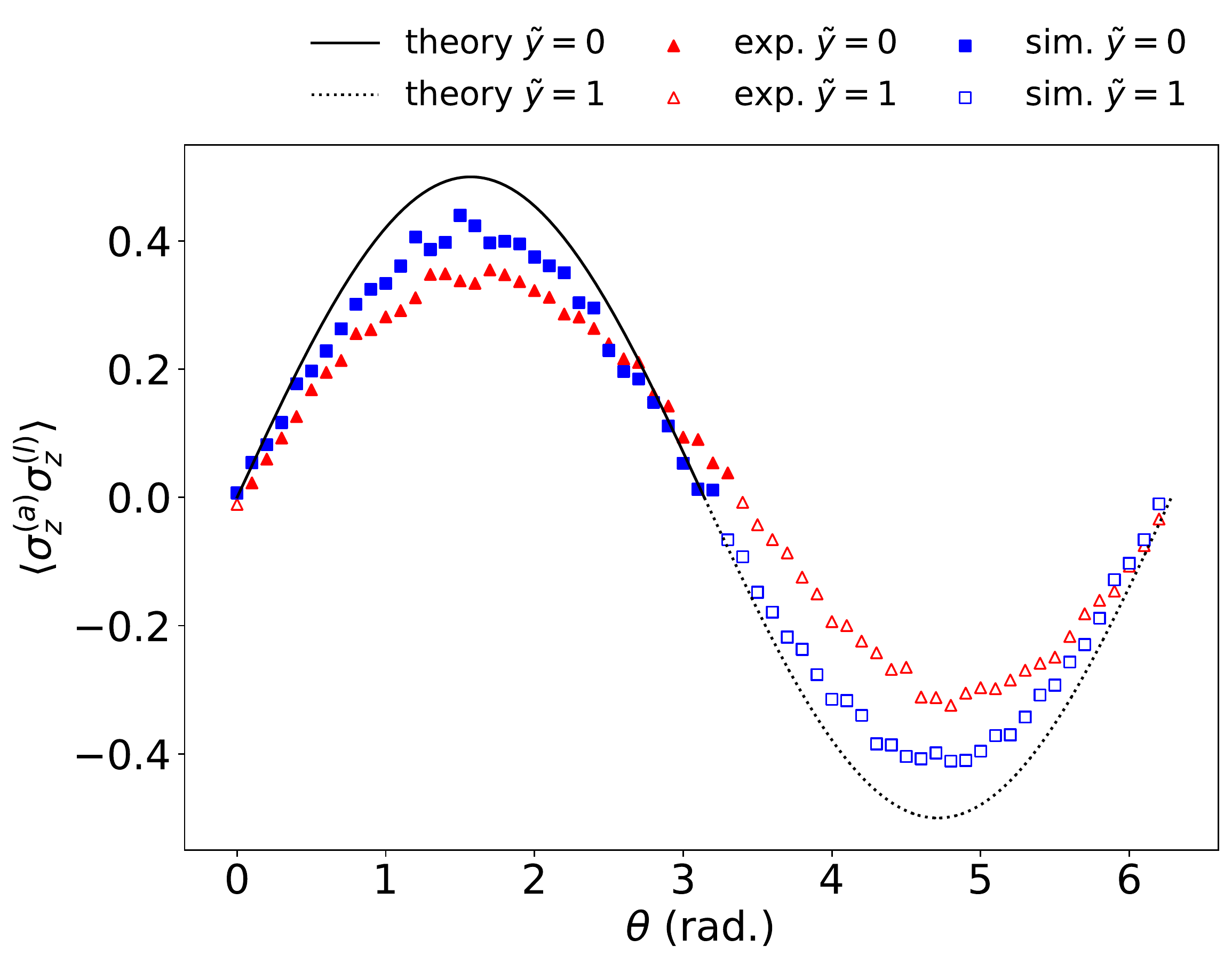}
	\caption{Classification of the toy problem outlined in Eqs.~(\ref{eqn:data_example}) and~(\ref{eqn:example_classification_statistics}) of the main manuscript vs. $\theta$. The experiment is performed on the ibmq\_ourense with date 2019-09-29, and its result (red triangles) is compared to simulation result (blue squares) obtained using device parameters listed in Supplementary Table~\ref{table:device_data} and to the theoretical values (black line).}
	\label{fig:4}
\end{figure}

In order to automate this procedure and ensure the same quality of each experiment, we have developed a scheduler based on a Python library called \texttt{Dask}~\cite{SUPPL_Dask}.

\begin{figure}[h]
    \centering
    \subfloat[\texttt{ibmqx4} -- 2019-03-24 10:27:57.008000 UTC]{
    \includegraphics[scale=0.35]{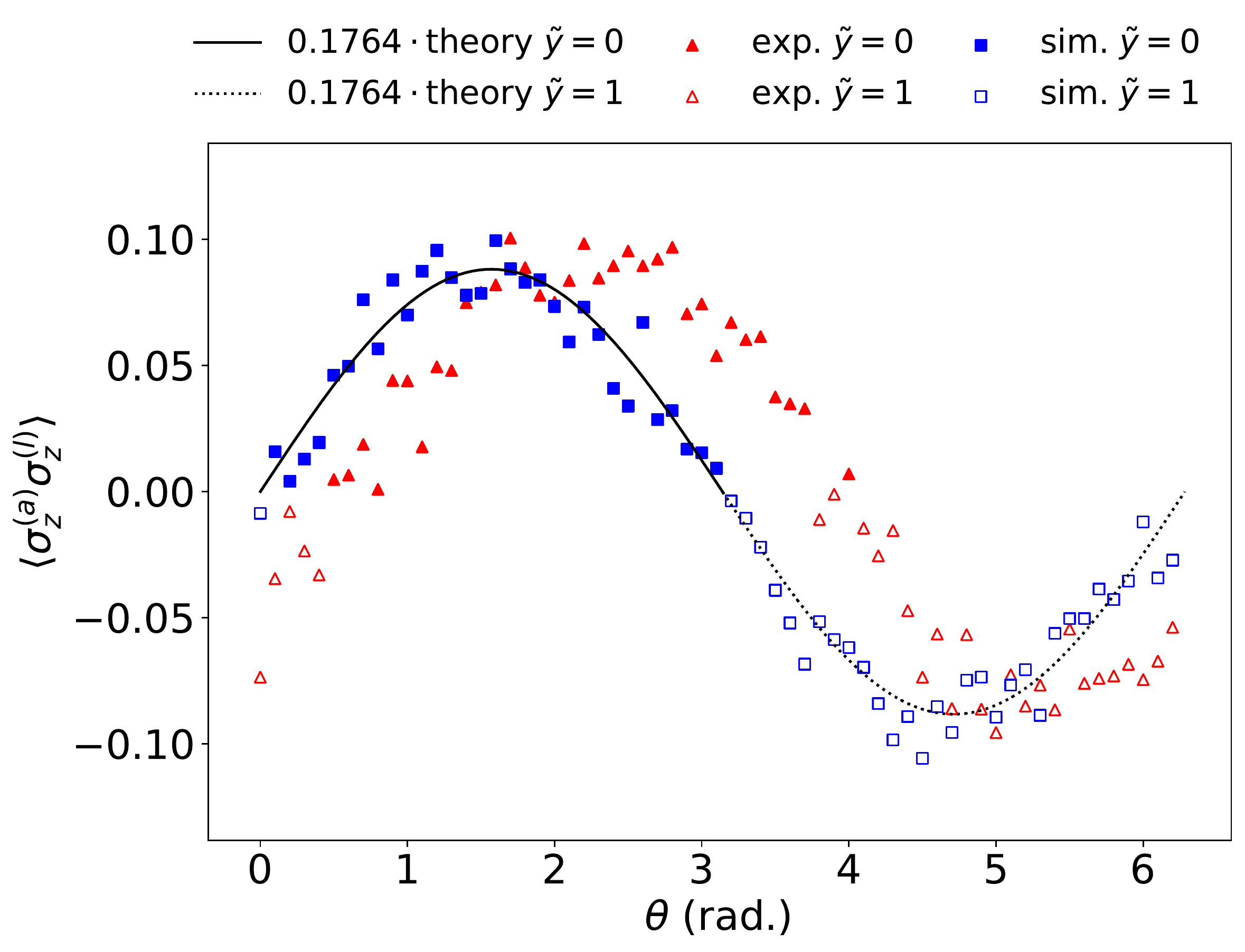}
    }
    \subfloat[\texttt{ibmq\_vigo} -- 2019-09-29 19:17:34.544799 UTC]{
    \includegraphics[scale=0.35]{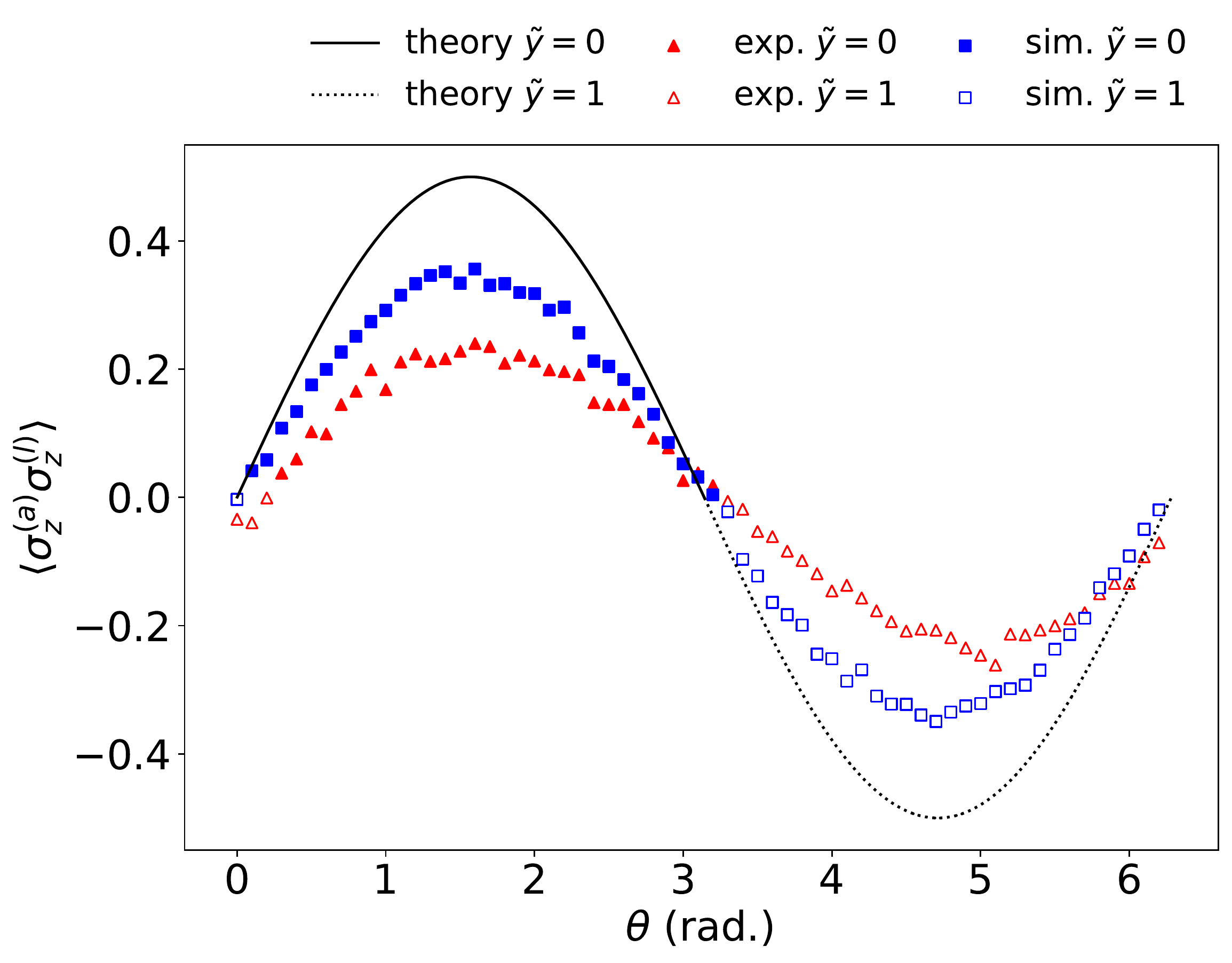}
    }
    \\
    \subfloat[\texttt{ibmqx2} -- 2019-09-29 19:36:19.920191 UTC]{
    \includegraphics[scale=0.35]{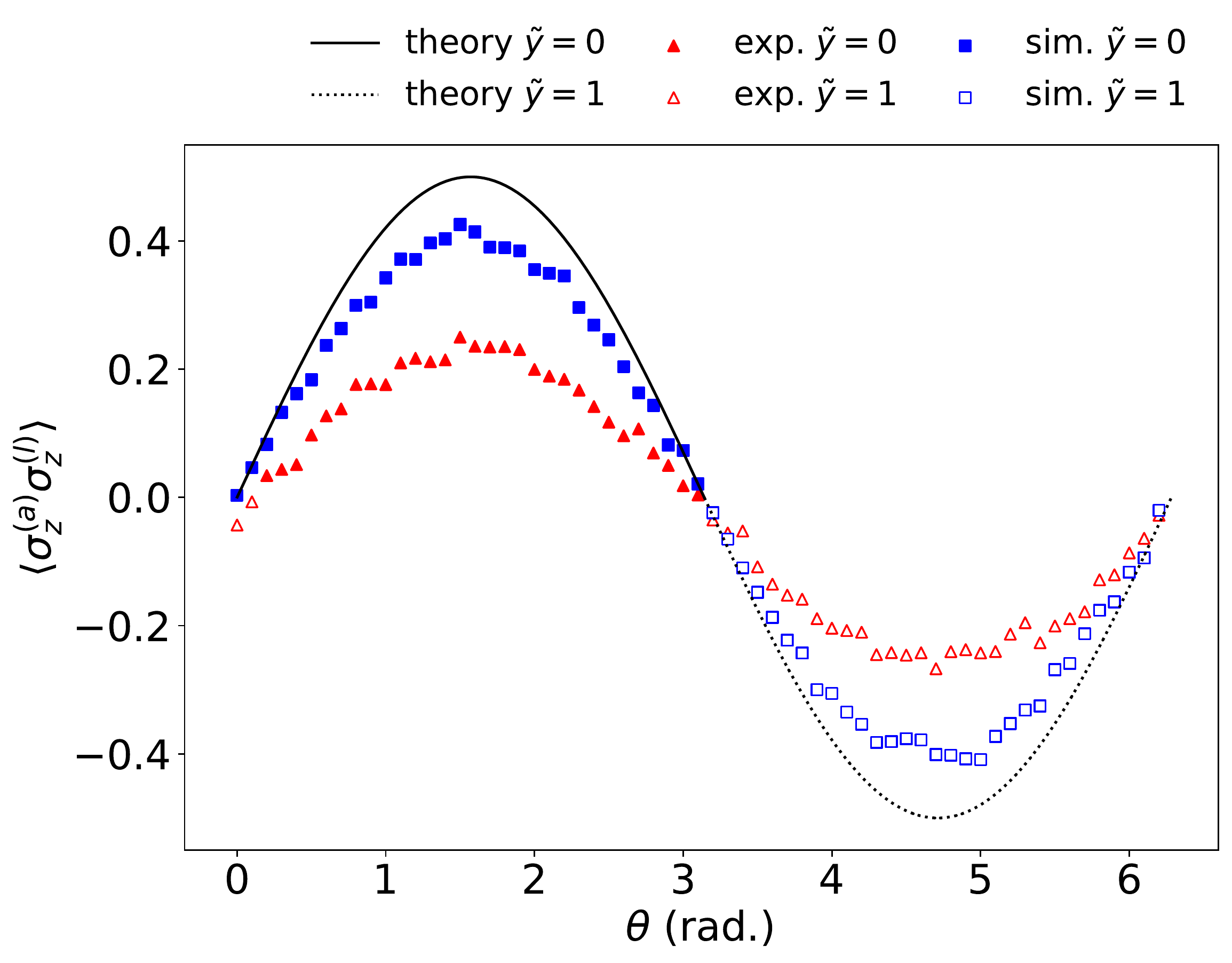}
    }
    \subfloat[\texttt{ibmq\_ourense} -- 2019-12-09 08:33:45.153361 UTC]{
    \includegraphics[scale=0.35]{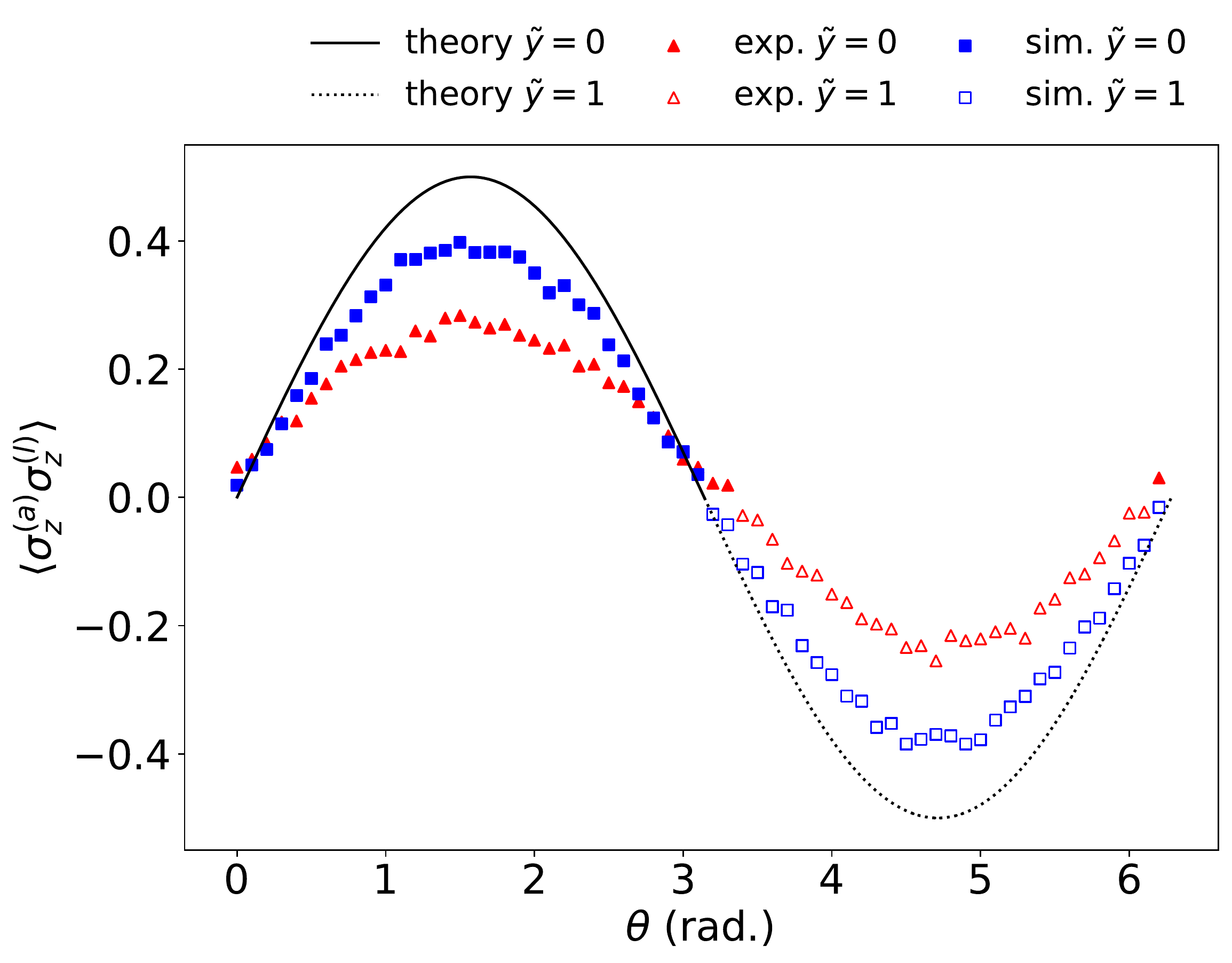}
    }
    \caption{Experimental results (triangle) from four different backends provided by IBM, performed at various times, and corresponding simulation results (square). Theoretical results are also plotted (solid and dotted lines). Note that the theoretical result in (a) is scaled by a factor of about 0.18 to improve the visibility when comparing to simulation and experimental results. Difference between simulation and experimental results in each plot can be attributed to various cross-talk effects, time-dependent noise, and non-Markovian noise.}
    \label{fig:gallery}
\end{figure}

For an analysis of the fitness of the noise simulation, we want to quantitatively assert the differences of both experimental and simulation results compared to the theoretical result. For this we define a reference function with parameters for the amplitude, phase shift and ordinate shift
$
f(a, \vartheta, w_2)(\theta) = \langle \sigma_z^{(a)}\sigma_z^{(y)}\rangle = a (\sin^2 {\left( \frac{\theta + \vartheta}{2} + \frac{\pi}{4} \right)} - w_2)
$
and fit this model to the data. By using the standard \texttt{scipy.optimize} we find for the benchmark (theory) $a \approx 9.99999993\mathrm{e}{-01}$, $\vartheta \approx -6.91619552\mathrm{e}{-09}$, $w_2 \approx 4.99999993\mathrm{e}{-01}$ which of course was expected. For the simulation and experiment we get, respectively,
\begin{align*}
    a \approx&\ 0.8213, &\vartheta &\approx -9/104329 \pi, &w_2 &\approx 0.50232985 \\
    a \approx&\ 0.6515, &\vartheta &\approx 2/51 \pi, &w_2 &\approx 0.5414.
\end{align*}
The discrepancies between the simulation and experimental results make apparent that the noise model considered in the simulation does not fully describe the experiment. From the thorough analysis of the applied noise model in Supplementary Note~\ref{subsec:simulation_noise}, we find that \texttt{qiskit-aer} error model does not include any ansatz for non-Markovian (see e.g. Ref.~\cite{SUPPL_1908.09855} for a definition) noise. While the dampening factor can be partially attributed to Pauli errors that do not commute with the measurement operators (as explained in the main manuscript), all other effects may be accounted for by various cross-talk effects, time-dependent noise, non-Markovian noise, and underestimation of the depolarizing, readout error, and relaxation rates. Rigorous device analysis for fully characterizing the device-dependent noise model is beyond the scope of our work and a field of research by itself. However, we invite all interested readers to dive into our data and code (see~\cite{SUPPL_SupplementaryRepository}) and observe how results differ during the various stages of development of the quantum devices. We want to conclude this section with a small gallery of experimental results obtained from various cloud quantum computers in different times in Fig.~\ref{fig:gallery}.

\subsection{Data and Code}
\label{sec:data}

The data can be found on GitHub~\cite{SUPPL_SupplementaryRepository}. The folder \verb|/experiment_results| (where \verb|/| means the root of the repository) holds all data referenced in the paper and the supplemental information. Important to note, all experiments with the ending \verb|_archive| are those experiments which do not have a matching noise simulation, i.e., the parameters used in the noise simulation are artificial as they are not directly collected from the actual quantum device at the time of the experiment. The others do. How to read the data is explained in the \verb|ReadMe.md| file accompanying the repository. We used the following data in the main manuscript:
\begin{itemize}
\item For the \textit{swap-test} classifier on the 2019-09-29 on \texttt{ibmq\_ourense}:
\texttt{exp\_sim\_regular\_20190929T114806Z.py}

\item For the \textit{swap-test} classifier on the 2019-03-24 on \texttt{ibmqx4}:
\texttt{exp\_sim\_regular\_noise\_job\_20190324T102757Z\_archive.py}
\item For the \textit{swap-test} classifier on the 2019-09-29 on \texttt{ibmq\_vigo}: \texttt{exp\_sim\_regular\_20190929T191722Z.py}
\item For the \textit{swap-test} classifier on the 2019-09-29 on \texttt{ibmqx2}: \texttt{exp\_sim\_regular\_20190929T193610Z.py}
\item For the \textit{swap-test} classifier on the 2019-12-09 on \texttt{ibmq\_ourense}: \texttt{exp\_sim\_regular\_20191209T083338Z.py}

\end{itemize}
There are many more experiments to be found that show the extent and history of our experiments.

\section{Supplementary Note: Listings}

\subsection{Circuit Factory Python Code}\label{fig:circuit_python_code}

The factory creating the \textit{swap-test} classifier is shown below. The function `compute\_rotation' computes an angle for a $Y$-rotation for preparing the index to a state that corresponds to the weights $w_1$ and $w_2$. The special gate \texttt{Ourense\_Fredkin} is a regular Fredkin (controlled-swap) gate but with a swap between the registers \texttt{qb\_in} and \texttt{qb\_d} ($q_2$ and $q_1$, respectively).

\begin{lstlisting}[language=Python]
import math
from typing import Optional, List

import qiskit
import qiskit.extensions.standard
from qiskit import QuantumCircuit, QuantumRegister, ClassicalRegister
from qiskit.extensions.standard.barrier import barrier

def create_swap_test_circuit_ourense(index_state, theta):
    # type: (List[float], float, Optional[dict]) -> QuantumCircuit
    """

    :param index_state:
    :param theta:
    :return:
    """
    use_barriers = kwargs.get('use_barriers', False)
    readout_swap = kwargs.get('readout_swap', None)

    q = QuantumRegister(5, "q")
    qb_a, qb_d, qb_in, qb_m, qb_l = (q[0], q[1], q[2], q[3], q[4])
    c = ClassicalRegister(2, "c")
    qc = QuantumCircuit(q, c, name="swap_test_ourense")

    # Index on q_0
    alpha_y, _ = compute_rotation(index_state)
    ry(qc, -alpha_y, qb_m).inverse()

    # Conditionally exite x_1 on data q_2 (center!)
    qc.h(qb_d)
    qc.rz(math.pi, qb_d).inverse()
    qc.s( qb_d)
    qc.cz(qb_m, qb_d)

    # Label y_1
    qc.cx(qb_m, qb_l)

    # Unknown data
    qc.rx(theta, qb_in)

    # Swap Test itself:
    # Hadamard on ancilla
    qc.h(qb_a)

    # c-SWAP!!!
    qc.append(Ourense_Fredkin(), [qb_a, qb_in, qb_d], [])

    # Hadamard on ancilla
    qc.h(qb_a)

    barrier(qc)
    qiskit.circuit.measure.measure(qc, qb_a, c[0])
    qiskit.circuit.measure.measure(qc, qb_l, c[1])

    return qc
\end{lstlisting}

\subsection{Expectation Value Python Code}\label{fig:expectation_value_python_code}

The code to calculate the two-qubit expectation value $\langle \sigma_z^{(a)} \sigma_z^{(l)} \rangle$ is shown below. 
\begin{lstlisting}[language=Python]
def extract_classification(counts):
    # type: (Dict[str, int]) -> float
    shots = sum(counts.values())
    return (counts.get('00', 0) - counts.get('01', 0) - \
            counts.get('10', 0) + counts.get('11', 0)) / float(shots)
\end{lstlisting}

\section{Supplementary Note: Circuits}
    \begin{figure}[H]
        \centering
        \includegraphics[width=1\columnwidth]{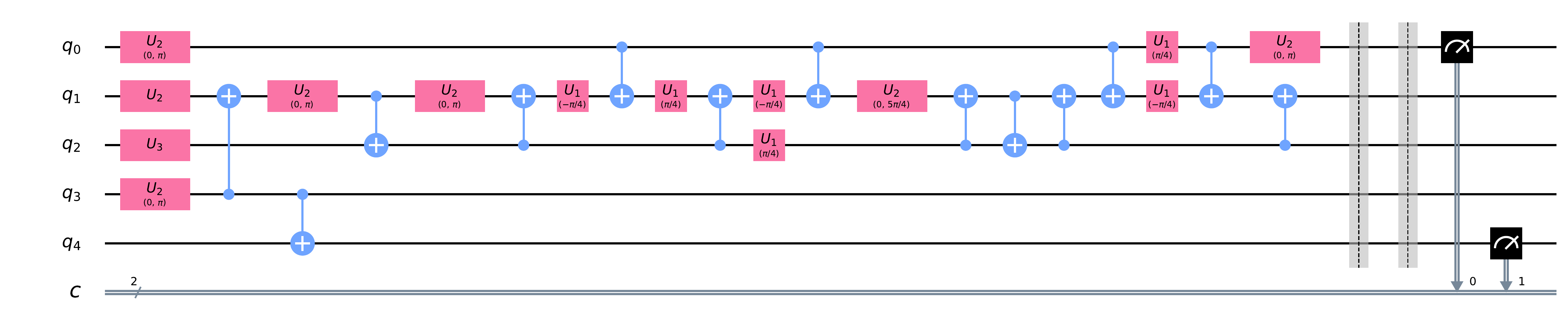}
        \caption{The transpiled circuit of the \textit{swap-test} classifier on \texttt{ibmq\_ourense}.\label{fig:transpiled_swaptest_circuit}}
        \label{}
    \end{figure}

	\begin{figure}[H]
		\centering
        \includegraphics[width=1\columnwidth]{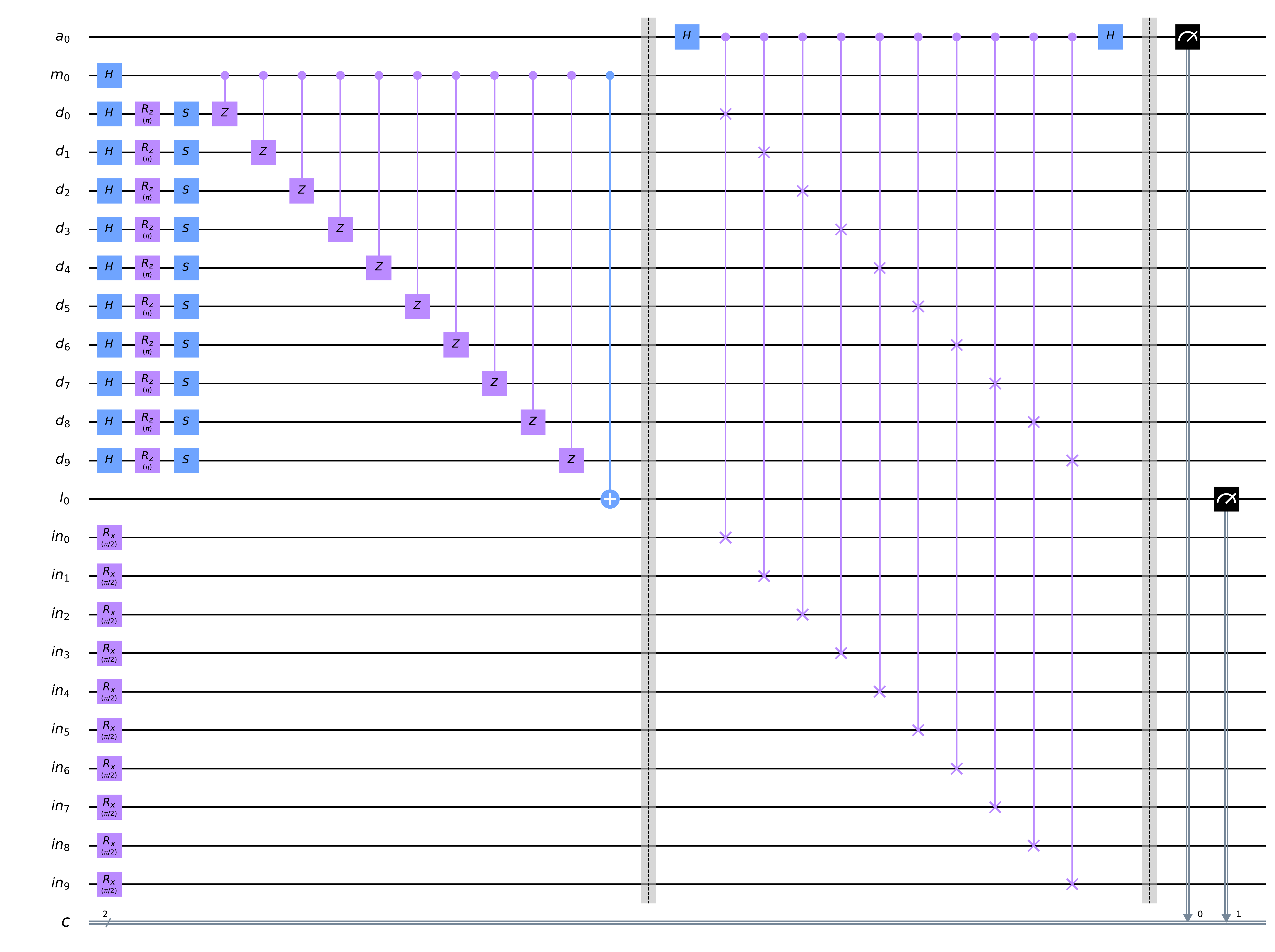}
		\caption{The high-level (non-compiled) circuit for 10-copies \textit{swap-test} classifier. \label{fig:TenCopiesHighLevelCircuit}}
	\end{figure}

\end{document}

%% file: circuits.tex
\newcommand{\controlpartial}{\scalebox{0.6}{\rotatebox{135}{\LEFTcircle}}}
\newcommand{\qwvdots}[1][-1]{\ar @{.} [#1,0]}
\newcommand{\qwcdots}{\push{\ \cdots\ }}
\newcommand{\ctrlp}[1]{\controlpartial \qwx[#1] \qw}
\newcommand{\ctrlpvdots}[1]{\controlpartial \qwvdots[#1] \qw}

\newcommand{\qtmlhadamard}{
\Qcircuit @C=1.0em @R=1.0em {
\lstick{\ket{a} = \ket{0}}    & \qw    &  \multigate{3}{\mathcal{U}\ket{0} = \ket{\Psi}} & \gate{H}  & \meter & \cw \cwx[3] & \cw \\
\lstick{\ket{m} = \ket{0}}    & {/}\qw &  \ghost{\mathcal{U}\ket{0} = \ket{\Psi}}        & \qw       & \qw    & \qw         & \qw \\
\lstick{\ket{d} = \ket{0}}    & {/}\qw &  \ghost{\mathcal{U}\ket{0} = \ket{\Psi}}        & \qw       & \qw    & \qw         & \qw \\
\lstick{\ket{l} = \ket{0}}    & \qw    &  \ghost{\mathcal{U}\ket{0} = \ket{\Psi}}        & \qw       & \qw    & \meter      & \cw
}
}

\newcommand{\quantumonly} {
\Qcircuit @C=0.8em @R=0.5em @!R {
\lstick{a: \ket{0}} & \qw & \qw & \qw & \qw & \qw & \qw & \qw & \qw & \qw & \qw & \qw & \qw & \qw & \gate{H} & \ctrl{2} & \qw & \gate{H} & \qw & \meter & \cw & \cw & \cw\\
\lstick{m: \ket{0}} & \qw & \qw & \qw & \qw & \gate{H} & \qw & \qw & \ctrlo{3} & \ctrlo{4} & \qw & \qw & \ctrl{5} & \ctrl{6} & \qw & \qw & \qw & \qw & \qw & \qw & \qw & \qw & \qw\\
\lstick{d: \ket{0}} & {/}\qw & \qw & \qw & \qw & \qw & \qw & \qw & \qswap \qwx[2] & \qw & \qw & \qw & \qswap \qwx[4] & \qw & \qw & \qswap & \qw & \qw & \qw & \qw & \qw & \qw & \qw\\
\lstick{l: \ket{0}} & \qw & \qw & \qw & \qw & \qw & \qw & \qw & \qw & \qswap \qwx[2] & \qw & \qw & \qw & \qswap \qwx[4] & \qw & \qw & \qw & \qw & \qw & \qw & \meter & \cw & \cw\\
\lstick{x_{0}: \ket{0}} & {/}\qw & \gate{U_\Phi(x_0)} & \qw & \qw & \qw & \qw & \qw & \qswap & \qw & \qw & \qw & \qw & \qw & \qw & \qw & \qw & \qw & \qw & \qw & \qw & \qw & \qw\\
\lstick{l_{0}: \ket{0}} & \qw & \gate{X^{y_0}} & \qw & \qw & \qw & \qw & \qw & \qw & \qswap & \qw & \qw & \qw & \qw & \qw & \qw & \qw & \qw & \qw & \qw & \qw & \qw & \qw\\
\lstick{x_{1}: \ket{0}} & {/}\qw & \gate{U_\Phi(x_1)} & \qw & \qw & \qw & \qw & \qw & \qw & \qw & \qw & \qw & \qswap & \qw & \qw & \qw & \qw & \qw & \qw & \qw & \qw & \qw & \qw\\
\lstick{l_{1}: \ket{0}} & \qw & \gate{X^{y_1}} & \qw & \qw & \qw & \qw & \qw & \qw & \qw & \qw & \qw & \qw & \qswap & \qw & \qw & \qw & \qw & \qw & \qw & \qw & \qw & \qw\\
\lstick{in: \ket{0}} & {/}\qw & \gate{U_\Phi(\tilde{x})} & \qw & \qw & \qw & \qw & \qw & \qw & \qw & \qw & \qw & \qw & \qw & \qw & \qswap \qwx[-6] & \qw & \qw & \qw & \qw & \qw & \qw & \qw\\
}
}

\newcommand{\quantumOnlyFourTrain}{
\Qcircuit @C=0.5em @R=0.0em @!R {
    \lstick{a: \ket{0}} & \qw & \qw & \qw & \gate{H} & \qw & \qw & \qw & \qw & \qw & \qw & \qw & \qw & \qw & \qw & \qw & \qw & \qw & \ctrl{3} & \qw & \gate{H} & \qw & \meter &\cw& \cw\\
    \lstick{m_{0}: \ket{0}} & \qw & \qw & \qw & \gate{H} & \qw & \ctrlo{1} & \ctrlo{1} & \qw & \ctrlo{1} & \ctrlo{1} & \qw & \ctrl{1} & \ctrl{1} & \qw & \ctrl{1} & \ctrl{1} & \qw & \qw & \qw & \qw & \qw & \qw & \qw & \qw\\
    \lstick{m_{1}: \ket{0}} & \qw & \qw & \qw & \gate{H} & \qw & \ctrlo{3} & \ctrlo{4} & \qw & \ctrl{5} & \ctrl{6} & \qw & \ctrlo{7} & \ctrlo{8} & \qw & \ctrl{9} & \ctrl{10} & \qw & \qw & \qw & \qw & \qw & \qw & \qw & \qw\\
    \lstick{d: \ket{0}} & \qw & \qw & \qw & \qw & \qw & \qswap \qwx[2] & \qw & \qw & \qswap \qwx[4] & \qw & \qw & \qswap \qwx[6] & \qw & \qw & \qswap \qwx[8] & \qw & \qw & \qswap & \qw & \qw & \qw & \qw & \qw & \qw\\
    \lstick{l: \ket{0}} & \qw & \qw & \qw & \qw & \qw & \qw & \qswap \qwx[2] & \qw & \qw & \qswap \qwx[4] & \qw & \qw & \qswap \qwx[6] & \qw & \qw & \qswap \qwx[8] & \qw & \qw & \qw & \qw & \qw & \qw & \meter & \cw\\
    \lstick{x_1: \ket{0}} & \gate{X} & \gate{H} & \qw & \qw & \qw & \qswap & \qw & \qw & \qw & \qw & \qw & \qw & \qw & \qw & \qw & \qw & \qw & \qw & \qw & \qw & \qw & \qw & \qw & \qw\\
    \lstick{l_1: \ket{0}} & \gate{X} & \qw & \qw & \qw & \qw & \qw & \qswap & \qw & \qw & \qw & \qw & \qw & \qw & \qw & \qw & \qw & \qw & \qw & \qw & \qw & \qw & \qw & \qw & \qw\\
    \lstick{x_2: \ket{0}} & \gate{H} & \qw & \qw & \qw & \qw & \qw & \qw & \qw & \qswap & \qw & \qw & \qw & \qw & \qw & \qw & \qw & \qw & \qw & \qw & \qw & \qw & \qw & \qw & \qw\\
    \lstick{l_2: \ket{0}} & \qw & \qw & \qw & \qw & \qw & \qw & \qw & \qw & \qw & \qswap & \qw & \qw & \qw & \qw & \qw & \qw & \qw & \qw & \qw & \qw & \qw & \qw & \qw & \qw\\
    \lstick{x_3: \ket{0}} & \gate{X} & \gate{H} & \qw & \qw & \qw & \qw & \qw & \qw & \qw & \qw & \qw & \qswap & \qw & \qw & \qw & \qw & \qw & \qw & \qw & \qw & \qw & \qw & \qw & \qw\\
    \lstick{l_3: \ket{0}} & \gate{X} & \qw & \qw & \qw & \qw & \qw & \qw & \qw & \qw & \qw & \qw & \qw & \qswap & \qw & \qw & \qw & \qw & \qw & \qw & \qw & \qw & \qw & \qw & \qw\\
    \lstick{x_4: \ket{0}} & \gate{H} & \qw & \qw & \qw & \qw & \qw & \qw & \qw & \qw & \qw & \qw & \qw & \qw & \qw & \qswap & \qw & \qw & \qw & \qw & \qw & \qw & \qw & \qw & \qw\\
    \lstick{l_4: \ket{0}} & \qw & \qw & \qw & \qw & \qw & \qw & \qw & \qw & \qw & \qw & \qw & \qw & \qw & \qw & \qw & \qswap & \qw & \qw & \qw & \qw & \qw & \qw & \qw & \qw\\
    \lstick{in: \ket{0}} & \gate{H} & \qw & \qw & \qw & \qw & \qw & \qw & \qw & \qw & \qw & \qw & \qw & \qw & \qw & \qw & \qw & \qw & \qswap \qwx[-10] & \qw & \qw & \qw & \qw & \qw & \qw\\
}
}
\newcommand{\quantumOnlyTwoDimData}{
\Qcircuit @C=0.5em @R=0.0em @!R {
    \lstick{a: \ket{0}}        & \qw        & \qw        & \qw   & \qw        & \qw   & \qw         & \qw         & \qw         & \qw        & \qw        & \qw        & \gate{H} & \ctrl{12} & \gate{H} & \qw & \qw  & \meter   &\cw       & \cw\\  
	\lstick{m \ket{0}}         & \qw        & \qw        & \qw   & \gate{H}   & \qw   & \ctrlo{4}   & \ctrlo{5}   & \ctrlo{6}   & \ctrl{7}   & \ctrl{8}   & \ctrl{9}   & \qw      & \qw      & \qw      & \qw & \qw  & \qw      & \qw      & \qw\\  
	\lstick{d_{1}: \ket{0}}    & \qw        & \qw        & \qw   & \qw        & \qw   & \qswap      & \qw         & \qw         & \qswap     & \qw        & \qw        & \qw      & \qswap   & \qw      & \qw & \qw  & \qw      & \qw      & \qw\\  
	\lstick{d_{2}: \ket{0}}    & \qw        & \qw        & \qw   & \qw        & \qw   & \qw         & \qswap      & \qw         & \qw        & \qswap     & \qw        & \qw      & \qw      & \qw      & \qw & \qw  & \qw      & \qw      & \qw\\  
	\lstick{l: \ket{0}}        & \qw        & \qw        & \qw   & \qw        & \qw   & \qw         & \qw         & \qswap      & \qw        & \qw        & \qswap     & \qw      & \qw      & \qw      & \qw & \qw  & \qw      & \meter   & \cw\\  
	\lstick{x_{11}: \ket{0}}   & \gate{X}   & \gate{H}   & \qw   & \qw        & \qw   & \qswap      & \qw         & \qw         & \qw        & \qw        & \qw        & \qw      & \qw      & \qw      & \qw & \qw  & \qw      & \qw      & \qw\\  
	\lstick{x_{12}: \ket{0}}   & \gate{X}   & \gate{H}   & \qw   & \qw        & \qw   & \qw         & \qswap      & \qw         & \qw        & \qw        & \qw        & \qw      & \qw      & \qw      & \qw & \qw  & \qw      & \qw      & \qw\\  
	\lstick{l_1: \ket{0}}      & \gate{X}   & \qw        & \qw   & \qw        & \qw   & \qw         & \qw         & \qswap      & \qw        & \qw        & \qw        & \qw      & \qw      & \qw      & \qw & \qw  & \qw      & \qw      & \qw\\  
	\lstick{x_{21}: \ket{0}}   & \gate{H}   & \qw        & \qw   & \qw        & \qw   & \qw         & \qw         & \qw         & \qswap     & \qw        & \qw        & \qw      & \qw      & \qw      & \qw & \qw  & \qw      & \qw      & \qw\\  
	\lstick{x_{22}: \ket{0}}   & \gate{X}   & \gate{H}   & \qw   & \qw        & \qw   & \qw         & \qw         & \qw         & \qw        & \qswap     & \qw        & \qw      & \qw      & \qw      & \qw & \qw  & \qw      & \qw      & \qw\\  
	\lstick{l_2: \ket{0}}      & \qw        & \qw        & \qw   & \qw        & \qw   & \qw         & \qw         & \qw         & \qw        & \qw        & \qswap     & \qw      & \qw      & \qw      & \qw & \qw  & \qw      & \qw      & \qw\\  
	\lstick{in_1: \ket{0}}     & \gate{H}   & \qw        & \qw   & \qw        & \qw   & \qw         & \qw         & \qw         & \qw        & \qw        & \qw        & \qw      & \qw      & \qw      & \qw & \qw  & \qw      & \qw      & \qw\\  
	\lstick{in_2: \ket{0}}     & \gate{H}   & \qw        & \qw   & \qw        & \qw   & \qw         & \qw         & \qw         & \qw        & \qw        & \qw        & \qw      & \qswap   & \qw      & \qw & \qw  & \qw      & \qw      & \qw\\  
}
}

\newcommand{\tildeU}{\tilde{\mathcal{U}}_\psi (D) }
\newcommand{\trainCompileQuantumInput} {
\Qcircuit @C=0.5em @R=0.0em @!R {
\lstick{a: \ket{0}}     & \qw    & \qw & \qw                    & \gate{H}                  & \qw  & \ctrl{2}        & \qw & \gate{H} & \qw & \meter & \cw \cwx[3] & \cw & \cw & \cw\\
\lstick{m: \ket{0}}     & {/}\qw & \qw & \multigate{2}{\tildeU} & \qw                       & \qw  & \qw             & \qw & \qw & \qw & \qw & \qw & \qw & \qw & \qw\\
\lstick{d: \ket{0}}     & {/}\qw & \qw & \ghost{{\tildeU}}      & \qw                       & \qw  & \qswap          & \qw & \qw & \qw & \qw & \qw & \qw & \qw & \qw\\
\lstick{l: \ket{0}}     & \qw    & \qw & \ghost{{\tildeU}}      & \qw                       & \qw  & \qw             & \qw & \qw & \qw & \qw & \meter & \cw & \cw & \cw\\
\lstick{in: \ket{0}}    & {/}\qw & \qw & \qw                    & \gate{U_\psi(\tilde{\x})} & \qw  & \qswap \qwx[-2] & \qw & \qw & \qw & \qw & \qw & \qw & \qw & \qw\\
}
}

\newcommand{\trainCompileQuantumInputTwoData} {
	\Qcircuit @C=0.5em @R=0.5em @!R {
\lstick{a: \ket{0}}        & \qw     & \qw  & \qw                     & \gate{H}                   & \qw   & \ctrl{2}         & \ctrl{3}         & \qw  & \gate{H}  & \qw  & \meter  & \cw \cwx[4]  & \cw  & \cw  & \cw\\ 
\lstick{m: \ket{0}}        & {/}\qw  & \qw  & \multigate{3}{\tildeU}  & \qw                        & \qw   & \qw              & \qw              & \qw  & \qw       & \qw  & \qw     & \qw          & \qw  & \qw  & \qw\\ 
\lstick{d_0: \ket{0}}      & {/}\qw  & \qw  & \ghost{{\tildeU}}       & \qw                        & \qw   & \qswap           & \qw              & \qw  & \qw       & \qw  & \qw     & \qw          & \qw  & \qw  & \qw\\ 
\lstick{d_1: \ket{0}}      & {/}\qw  & \qw  & \ghost{{\tildeU}}       & \qw                        & \qw   & \qw              & \qswap           & \qw  & \qw       & \qw  & \qw     & \qw          & \qw  & \qw  & \qw\\ 
\lstick{l: \ket{0}}        & \qw     & \qw  & \ghost{{\tildeU}}       & \qw                        & \qw   & \qw              & \qw              & \qw  & \qw       & \qw  & \qw     & \meter       & \cw  & \cw  & \cw\\ 
\lstick{in_0: \ket{0}}     & {/}\qw  & \qw  & \qw                     & \gate{U_\psi(\tilde{\x})}  & \qw   & \qswap \qwx[-3]  & \qw              & \qw  & \qw       & \qw  & \qw     & \qw          & \qw  & \qw  & \qw\\ 
\lstick{in_1: \ket{0}}     & {/}\qw  & \qw  & \qw                     & \gate{U_\psi(\tilde{\x})}  & \qw   & \qw              & \qswap \qwx[-3]  & \qw  & \qw       & \qw  & \qw     & \qw          & \qw  & \qw  & \qw\\ 
}
}

\newcommand{\NisqExperimentCircuit} {
	\Qcircuit @C=0.5em @R=0.0em @!R {
	\lstick{a: \ket{0}}   & \gate{H}            & \qw                        & \qw       & \qw      & \qw  & \qw       & \qw  & \ctrl{2}         & \qw  & \gate{H}  & \qw  & \meter  & \cw     & \cw  & \cw  & \cw\\ 
	\lstick{m: \ket{0}}   & \gate{R^\dagger_y(\alpha)}            & \qw                        & \qw       & \ctrl{1} & \qw  & \ctrl{2}  & \qw  & \qw              & \qw  & \qw       & \qw  & \qw     & \qw     & \qw  & \qw  & \qw\\ 
	\lstick{d: \ket{0}}   & \gate{H}            & \gate{R^\dagger_z(\pi)}    & \gate{S}  & \gate{Z} & \qw  & \qw       & \qw  & \qswap           & \qw  & \qw       & \qw  & \qw     & \qw     & \qw  & \qw  & \qw\\ 
	\lstick{l: \ket{0}}   & \qw                 & \qw                        & \qw       & \qw      & \qw  & \targ     & \qw  & \qw              & \qw  & \qw       & \qw  & \qw     & \meter  & \cw  & \cw  & \cw\\ 
	\lstick{in: \ket{0}}  & \gate{R_x(\theta)}  & \qw                        & \qw       & \qw      & \qw  & \qw       & \qw  & \qswap \qwx[-2]  & \qw  & \qw       & \qw  & \qw     & \qw     & \qw  & \qw  & \qw\\ 
	}
}

\newcommand{\NisqExperimentCircuitHadamard} {
	\Qcircuit @C=0.5em @R=0.0em @!R {
	\lstick{a: \ket{0}}    & \gate{H}                      & \ctrlo{2}   & \ctrlo{2}                   & \ctrlo{2}  & \ctrlo{2} & \qw   & \qw        & \qw   & \ctrl{2}           & \qw   & \gate{H}   & \qw   & \meter   & \cw      & \cw   & \cw   & \cw\\  
	\lstick{m: \ket{0}}    & \gate{R^\dagger_y(\alpha)}    & \qw         & \qw                         & \qw        & \ctrl{1}  & \qw   & \ctrl{2}   & \qw   & \qw                & \qw   & \qw        & \qw   & \qw      & \qw      & \qw   & \qw   & \qw\\  
	\lstick{d: \ket{0}}    & \qw                           & \gate{H}    & \gate{R^\dagger_z(\pi)}     & \gate{S}   & \gate{Z}  & \qw   & \qw        & \qw   & \gate{R_x(\theta)} & \qw   & \qw        & \qw   & \qw      & \qw      & \qw   & \qw   & \qw\\  
	\lstick{l: \ket{0}}    & \qw                           & \qw         & \qw                         & \qw        & \qw       & \qw   & \targ      & \qw   & \qw                & \qw   & \qw        & \qw   & \qw      & \meter   & \cw   & \cw   & \cw
	}
}

\newcommand{\multicopyclassifier} {
	\Qcircuit @C=0.8em @R=0.5em @!R {
\lstick{a: \ket{0}}           & \qw        & \qw                        & \qw      & \qw              & \qw             & \qw             & \qw             & \qw                & \qw                 & \gate{H}     & \ctrl{2}            & \ctrl{3}          & \gate{H}  & \meter     & \cw        & \cw     & \cw\\    
\lstick{m: \ket{0}}           & \qw        & \qw                        & \gate{H} & \ctrlo{1}        & \ctrlo{2}       & \ctrlo{3}       & \ctrl{1}        & \ctrl{2}           & \ctrl{3}            & \qw          & \qw                 & \qw               & \qw       & \qw        & \qw        & \qw     & \qw\\    
\lstick{d_0: \ket{0}}         & {/}\qw     & \qw                        & \qw      & \qswap \qwx[3]   & \qw             & \qw             & \qswap \qwx[3]  & \qw                & \qw                 & \qw          & \qswap \qwx[3]      & \qw               & \qw       & \qw        & \qw        & \qw     & \qw\\    
\lstick{d_1: \ket{0}}         & {/}\qw     & \qw                        & \qw      & \qw              & \qswap \qwx[2]  & \qw             & \qw             & \qswap \qwx[2]     & \qw                 & \qw          & \qw                 & \qswap \qwx[2]    & \qw       & \qw        & \qw        & \qw     & \qw\\    
\lstick{l: \ket{0}}           & \qw        & \qw                        & \qw      & \qw              & \qw             & \qswap \qwx[3]  & \qw             & \qw                & \qswap \qwx[5]      & \qw          & \qw                 & \qw               & \qw       & \qw        & \qw        & \qw     & \qw\\    
\lstick{s: \ket{0}}           & \qw        & \qw                        & \gate{H} & \ctrlo{1}        & \ctrl{1}        & \qw             & \ctrlo{3}       & \ctrl{3}           & \qw                 & \qw          & \ctrlo{5}           & \ctrl{5}          & \qw       & \qw        & \meter     & \cw     & \cw\\    
\lstick{x_{0}: \ket{0}}       & {/}\qw     & \gate{U_\Phi(x_0)}         & \qw      & \qswap           & \qswap          & \qw             & \qw             & \qw                & \qw                 & \qw          & \qw                 & \qw               & \qw       & \qw        & \qw        & \qw     & \qw\\    
\lstick{l_{0}: \ket{0}}       & \qw        & \gate{X^{y_0}}             & \qw      & \qw              & \qw             & \qswap          & \qw             & \qw                & \qw                 & \qw          & \qw                 & \qw               & \qw       & \qw        & \qw        & \qw     & \qw\\    
\lstick{x_{1}: \ket{0}}       & {/}\qw     & \gate{U_\Phi(x_1)}         & \qw      & \qw              & \qw             & \qw             & \qswap          & \qswap             & \qw                 & \qw          & \qw                 & \qw               & \qw       & \qw        & \qw        & \qw     & \qw\\    
\lstick{l_{1}: \ket{0}}       & \qw        & \gate{X^{y_1}}             & \qw      & \qw              & \qw             & \qw             & \qw             & \qw                & \qswap              & \qw          & \qw                 & \qw               & \qw       & \qw        & \qw        & \qw     & \qw\\    
\lstick{in: \ket{0}}          & {/}\qw     & \gate{U_\Phi(\tilde{x})}   & \qw      & \qw              & \qw             & \qw             & \qw             & \qw                & \qw                 & \qw          & \qswap              & \qswap            & \qw       & \qw        & \qw        & \qw     & \qw\\ 
}
}

\newcommand{\ibmqxfourimplementation}{
\Qcircuit @C=0.5em @R=0.0em @!R {
	\lstick{q_{0}: \ket{0}}  & \qw              & \qw               & \gate{R_y(\alpha)}  & \ctrl{2}     & \qw              & \ctrl{1}  & \qw  & \qw           & \qw           & \qw  & \qw\\ 
	\lstick{q_{1}: \ket{0}}  & \qw              & \qw               & \qw       & \qw          & \qw              & \targ     & \qw                       & \meter        & \qw           & \qw  & \qw\\ 
	\lstick{q_{2}: \ket{0}}  & \gate{H}         & \gate{R^\dagger_z(\pi)}  & \gate{S}  & \gate{Z}\qw  & \qswap           & \qw       & \qw                       & \qw           & \qw           & \qw  & \qw\\ 
	\lstick{q_{3}: \ket{0}}  & \gate{R_x(\theta)}  & \qw               & \qw       & \qw          & \qswap \qwx[-1]  & \qw       & \qw                       & \qw           & \qw           & \qw  & \qw\\ 
	\lstick{q_{4}: \ket{0}}  & \gate{H}         & \qw               & \qw       & \qw          & \ctrl{-2}        & \gate{H}  & \qw                       & \qw           & \meter        & \qw  & \qw\\ 
	\lstick{c_{0}: 0}        & \cw              & \cw               & \cw       & \cw          & \cw              & \cw       & \cw                       & \cw           & \cw \cwx[-1]  & \cw  & \cw\\ 
	\lstick{c_{1}: 0}        & \cw              & \cw               & \cw       & \cw          & \cw              & \cw       & \cw                       & \cw \cwx[-5]  & \cw           & \cw  & \cw\\ 
	 }
}

\newcommand{\ibmqxfourimplementationCompiled}{
\Qcircuit @C=0.5em @R=0.0em @!R {
\lstick{q_{0}: \ket{0}} & \gate{U_1(\pi)} & \targ & \qw & \qw & \qw & \qw & \qw & \qw & \qw & \qw & \qw & \qw & \qw & \qw & \qw & \qw & \targ & \gate{U_2\left(0,\pi\right)} & \qw & \qw & \qw & \qw & \qw\\
\lstick{q_{1}: \ket{0}} & \gate{U_2\left(0,\pi\right)} & \qw & \qw & \qw & \qw & \qw & \qw & \qw & \qw & \qw & \qw & \qw & \qw & \qw & \qw & \qw & \ctrl{-1} & \gate{U_2\left(0,\pi\right)} & \qw & \meter & \qw & \qw & \qw\\
\lstick{q_{2}: \ket{0}} & \gate{U_3(1.6,-1.6,3.1)} & \ctrl{-2} & \targ & \gate{U_2\left(0,\pi\right)} & \targ & \qw & \qw & \qw & \targ & \gate{U_2\left(\pi/4,\pi\right)} & \qw & \qw & \targ & \gate{U_1(-\pi/4)} & \targ & \qw & \targ & \qw & \qw & \qw & \qw & \qw & \qw\\
\lstick{q_{3}: \ket{0}} & \gate{U_3(\theta,-\pi/2,\pi/2)} & \qw & \ctrl{-1} & \qw & \ctrl{-1} & \gate{U_3(\pi/4,\pi/2,3*\pi/2)} & \ctrl{1} & \gate{U_3(-\pi/4,\pi/2,3*\pi/2)} & \ctrl{-1} & \gate{U_3(\pi/4,\pi/2,3*\pi/2)} & \ctrl{1} & \qw & \qw & \qw & \qw & \gate{U_3(-\pi/4,\pi/2,3*\pi/2)} & \ctrl{-1} & \qw & \qw & \qw & \qw & \qw & \qw\\
\lstick{q_{4}: \ket{0}} & \qw & \qw & \qw & \qw & \qw & \qw & \targ & \qw & \qw & \qw & \targ & \gate{U_2\left(0,\pi\right)} & \ctrl{-2} & \gate{U_1(\pi/4)} & \ctrl{-2} & \gate{U_2\left(0,\pi\right)} & \qw & \qw & \qw & \qw & \meter & \qw & \qw\\
\lstick{c_{0}: 0} & \cw & \cw & \cw & \cw & \cw & \cw & \cw & \cw & \cw & \cw & \cw & \cw & \cw & \cw & \cw & \cw & \cw & \cw & \cw & \cw & \cw \cwx[-1] & \cw & \cw\\
\lstick{c_{1}: 0} & \cw & \cw & \cw & \cw & \cw & \cw & \cw & \cw & \cw & \cw & \cw & \cw & \cw & \cw & \cw & \cw & \cw & \cw & \cw & \cw \cwx[-5] & \cw & \cw & \cw\\
	 }
}

\newcommand{\OriginalHighLevelCircuit}{
	\Qcircuit @C=0.5em @R=0.0em @!R {
		\lstick{q^d_{0}: \ket{0}}     & \qw          & \gate{U_3(\theta,-\frac{\pi}{2},\frac{\pi}{2})}     & \qw          & \gate{H}      & \gate{R^\dagger_z(\pi)}     & \gate{U_1(\frac{\pi}{2})}\qw     & \gate{H}     & \targ         & \gate{H}     & \qw          & \qw                      & \qw              & \qw              & \qw     & \qw\\    
		\lstick{q^a_{1}: \ket{0}}     & \gate{H}     & \ctrl{-1}                                           & \gate{X}     & \ctrl{-1}     & \ctrl{-1}                   & \ctrl{-1}                        & \qw          & \ctrl{1}      & \gate{X}     & \gate{H}     & \qw                          & \qw              & \meter           & \qw     & \qw\\    
		\lstick{q^m_{2}: \ket{0}}     & \gate{R_y(\alpha)}     & \qw                                                 & \qw          & \qw           & \qw                         & \qw                              & \qw          & \ctrl{-2}     & \ctrl{1}     & \qw          & \qw                          & \qw              & \qw              & \qw     & \qw\\    
		\lstick{q^l_{3}: \ket{0}}     & \qw          & \qw                                                 & \qw          & \qw           & \qw                         & \qw                              & \qw          & \qw           & \targ        & \qw          & \qw                          & \meter           & \qw              & \qw     & \qw\\    
		\lstick{c_{0}: 0}             & \cw          & \cw                                                 & \cw          & \cw           & \cw                         & \cw                              & \cw          & \cw           & \cw          & \cw          & \cw                          & \cw              & \cw \cwx[-3]     & \cw     & \cw\\    
		\lstick{c_{1}: 0}             & \cw          & \cw                                                 & \cw          & \cw           & \cw                         & \cw                              & \cw          & \cw           & \cw          & \cw          & \cw                          & \cw \cwx[-2]     & \cw              & \cw     & \cw\\    
	}
}

\newcommand{\OriginalLowLevelCircuit}{
	\Qcircuit @C=0.5em @R=0.0em @!R {
\lstick{q^d_{0}: \ket{0}}   & \gate{U_1(1.6)}               & \targ       & \gate{U_3(-\theta/2,0,0)}   & \targ       & \gate{U_3(1.6,-pi/2 + \theta/2,1.6)}   & \targ       & \gate{U_2\left(pi/4,pi\right)}   & \targ       & \gate{U_3(1.6,-3.1,0.79)}   & \targ       & \gate{U_1(1.6)}   & \targ       & \qw                & \targ       & \gate{U_1(-0.79)}   & \targ       & \gate{U_1(0.79 + 2*pi)}   & \targ       & \gate{U_1(-pi/4)}   & \targ       & \gate{U_1(pi/4)}   & \targ       & \gate{U_1(-pi/4)}   & \targ       & \gate{U_1(9*pi/4)}                & \qw         & \qw                             & \qw         & \qw               & \qw                           & \qw       & \qw            & \qw            & \qw   & \qw\\  
\lstick{q1a_{1}: \ket{0}}   & \gate{U_2\left(0,pi\right)}   & \ctrl{-1}   & \qw                         & \ctrl{-1}   & \gate{U_3(pi,0,pi)}                    & \ctrl{-1}   & \qw                              & \ctrl{-1}   & \gate{U_1(pi/2)}            & \ctrl{-1}   & \qw               & \ctrl{-1}   & \gate{U_1(0.79)}   & \ctrl{-1}   & \qw                 & \ctrl{-1}   & \qw                       & \qw         & \qw                 & \ctrl{-1}   & \qw                & \qw         & \qw                 & \ctrl{-1}   & \gate{U_2\left(0,pi\right)}       & \targ       & \gate{U_3(-pi/4,pi/2,3*pi/2)}   & \targ       & \gate{U_1(3.1)}   & \qw                           & \qw                        & \qw            & \meter         & \qw   & \qw\\  
\lstick{q^m_{2}: \ket{0}}   & \gate{U_2\left(0,pi\right)}   & \qw         & \qw                         & \qw         & \qw                                    & \qw         & \qw                              & \qw         & \qw                         & \qw         & \qw               & \qw         & \qw                & \qw         & \qw                 & \qw         & \qw                       & \ctrl{-2}   & \qw                 & \qw         & \qw                & \ctrl{-2}   & \qw                 & \qw         & \gate{U_2\left(0,5*pi/4\right)}   & \ctrl{-1}   & \gate{U_3(pi/4,pi/2,3*pi/2)}    & \ctrl{-1}   & \targ             & \gate{U_2\left(0,pi\right)}   & \qw                        & \qw            & \qw            & \qw   & \qw\\  
\lstick{q^l_{3}: \ket{0}}   & \gate{U_2\left(0,pi\right)}   & \qw         & \qw                         & \qw         & \qw                                    & \qw         & \qw                              & \qw         & \qw                         & \qw         & \qw               & \qw         & \qw                & \qw         & \qw                 & \qw         & \qw                       & \qw         & \qw                 & \qw         & \qw                & \qw         & \qw                 & \qw         & \qw                               & \qw         & \qw                             & \qw         & \ctrl{-1}         & \gate{U_2\left(0,pi\right)}   & \qw                        & \meter         & \qw            & \qw   & \qw\\  
\lstick{c_{0}: 0}         & \cw                           & \cw         & \cw                         & \cw         & \cw                                    & \cw         & \cw                              & \cw         & \cw                         & \cw         & \cw               & \cw         & \cw                & \cw         & \cw                 & \cw         & \cw                       & \cw         & \cw                 & \cw         & \cw                & \cw         & \cw                 & \cw         & \cw                               & \cw         & \cw                             & \cw         & \cw               & \cw                           & \cw                        & \cw            & \cw \cwx[-3]   & \cw   & \cw\\  
\lstick{c_{1}: 0}         & \cw                           & \cw         & \cw                         & \cw         & \cw                                    & \cw         & \cw                              & \cw         & \cw                         & \cw         & \cw               & \cw         & \cw                & \cw         & \cw                 & \cw         & \cw                       & \cw         & \cw                 & \cw         & \cw                & \cw         & \cw                 & \cw         & \cw                               & \cw         & \cw                             & \cw         & \cw               & \cw                           & \cw                        & \cw \cwx[-2]   & \cw            & \cw   & \cw\\  
	}
}

\newcommand{\SwapTestCompiledCircuitIbmQxFour}{
 \Qcircuit @C=0.5em @R=0.0em @!R {
	 	\lstick{q_{0}: \ket{0}} & \qw & \targ & \targ & \gate{U_2\left(0.0,\pi\right)} & \qw & \qw & \qw & \qw & \qw & \qw & \qw & \qw & \qw & \qw & \qw & \qw & \qw & \qw & \qw & \qw & \qw & \qw\\
	 	\lstick{q_{1}: \ket{0}} & \gate{U_2\left(0,\pi\right)} & \qw & \ctrl{-1} & \gate{U_2\left(0,\pi\right)} & \qw & \qw & \qw & \qw & \qw & \qw & \qw & \qw & \qw & \qw & \qw & \qw & \qw & \qw & \qw & \meter & \qw & \qw\\
	 	\lstick{q_{2}: \ket{0}} & \gate{U_3(\pi/2,-\pi/2,\pi)} & \ctrl{-2} & \qw & \targ & \gate{U_2\left(0,\pi\right)} & \targ & \qw & \qw & \qw & \targ & \qw & \qw & \gate{U_2\left(\pi/4,\pi\right)} & \targ & \gate{U_1(-\pi/4)} & \targ & \targ & \qw & \qw & \qw & \qw & \qw\\
	 	\lstick{q_{3}: \ket{0}} & \gate{U_3(\theta,-\pi/2,\pi/2)} & \qw & \qw & \ctrl{-1} & \qw & \ctrl{-1} & \gate{U_3(\pi/4,\pi/2,3\pi/2)} & \ctrl{1} & \gate{U_3(-\pi/4,\pi/2,3\pi/2)} & \ctrl{-1} & \gate{U_3(\pi/4,\pi/2,3\pi/2)} & \ctrl{1} & \gate{U_3(-\pi/4,\pi/2,3\pi/2)} & \qw & \qw & \qw & \ctrl{-1} & \qw & \qw & \qw & \qw & \qw\\
	 	\lstick{q_{4}: \ket{0}} & \qw & \qw & \qw & \qw & \qw & \qw & \qw & \targ & \qw & \qw & \qw & \targ & \gate{U_2\left(0,\pi\right)} & \ctrl{-2} & \gate{U_1(\pi/4)} & \ctrl{-2} & \gate{U_2\left(0,\pi\right)} & \qw & \meter & \qw & \qw & \qw\\
	 	\lstick{c_{0}: 0} & \cw & \cw & \cw & \cw & \cw & \cw & \cw & \cw & \cw & \cw & \cw & \cw & \cw & \cw & \cw & \cw & \cw & \cw & \cw \cwx[-1] & \cw & \cw & \cw\\
	 	\lstick{c_{1}: 0} & \cw & \cw & \cw & \cw & \cw & \cw & \cw & \cw & \cw & \cw & \cw & \cw & \cw & \cw & \cw & \cw & \cw & \cw & \cw & \cw \cwx[-5] & \cw & \cw\\
	 }
}

\newcommand{\SwapTestCompiledCircuitIbmQxTwo}{
 \Qcircuit @C=0.5em @R=0.0em @!R {
	 	\lstick{q_{0}: \ket{0}} & \gate{U_2\left(0,3.1\right)} & \ctrl{2} & \qw & \qw & \qw & \qw & \qw & \qw & \qw & \qw & \qw & \qw & \qw & \qw & \qw & \qw & \ctrl{1} & \qw & \qw & \qw & \qw & \qw\\
	 	\lstick{q_{1}: \ket{0}} & \qw & \qw & \qw & \qw & \qw & \qw & \qw & \qw & \qw & \qw & \qw & \qw & \qw & \qw & \qw & \qw & \targ & \qw & \qw & \meter & \qw & \qw\\
	 	\lstick{q_{2}: \ket{0}} & \gate{U_3(\pi/2,\pi/2,3\pi/2)} & \targ & \gate{U_2\left(0,\pi\right)} & \targ & \gate{U_2\left(0,\pi\right)} & \targ & \qw & \qw & \qw & \targ & \qw & \qw & \gate{U_2\left(\pi4,\pi\right)} & \targ & \gate{U_1(-\pi)} & \targ & \targ & \qw & \qw & \qw & \qw & \qw\\
	 	\lstick{q_{3}: \ket{0}} & \gate{U_3(\theta,-\pi/2,\pi/2)} & \qw & \qw & \ctrl{-1} & \qw & \ctrl{-1} & \gate{U_3(\pi/4,\pi/2,3\pi/2)} & \ctrl{1} & \gate{U_3(-\pi/4,\pi/2,3\pi/2)} & \ctrl{-1} & \gate{U_3(\pi/4,\pi/2,3\pi/2)} & \ctrl{1} & \gate{U_3(-\pi/4,\pi/2,3\pi/2)} & \qw & \qw & \qw & \ctrl{-1} & \qw & \qw & \qw & \qw & \qw\\
	 	\lstick{q_{4}: \ket{0}} & \qw & \qw & \qw & \qw & \qw & \qw & \qw & \targ & \qw & \qw & \qw & \targ & \gate{U_2\left(0,\pi\right)} & \ctrl{-2} & \gate{U_1(\pi/4)} & \ctrl{-2} & \gate{U_2\left(0,\pi\right)} & \qw & \meter & \qw & \qw & \qw\\
	 	\lstick{c_{0}: 0} & \cw & \cw & \cw & \cw & \cw & \cw & \cw & \cw & \cw & \cw & \cw & \cw & \cw & \cw & \cw & \cw & \cw & \cw & \cw \cwx[-1] & \cw & \cw & \cw\\
	 	\lstick{c_{1}: 0} & \cw & \cw & \cw & \cw & \cw & \cw & \cw & \cw & \cw & \cw & \cw & \cw & \cw & \cw & \cw & \cw & \cw & \cw & \cw & \cw \cwx[-5] & \cw & \cw\\
	 }
}

\newcommand{\ProductStateExampleCircuit}{
	\Qcircuit @C=0.5em @R=0.0em @!R {
		\lstick{a: \ket{0}}                 & \qw                   & \qw                              & \qw        & \qw            & \qw   & \qw        & \qw     & \qw                        & \qw    & \qw                        & \qw    & \qw            & \qw    & \qw                        & \qw   & \qw                        & \qw   & \qw            & \qw   & \gate{H}       & \ctrl{1}                        & \gate{H}       & \qw   & \qw                & \meter             & \qw       & \qw\\      
		\lstick{\tilde{x}: \ket{0}}         & \gate{R_x(\theta)}    & \qw                              & \qw        & \qw            & \qw   & \qw        & \qw     & \qw                        & \qw    & \qw                        & \qw    & \qw            & \qw    & \qw                        & \qw   & \qw                        & \qw   & \qw            & \qw   & \qw            & \qswap \qwx[1]                  & \qw            & \qw   & \qw                & \qw                & \qw       & \qw\\      
		\lstick{d: \ket{0}}                 & \qw                   & \qw                              & \qw        & \qw            & \qw   & \qw        & \qw     & \qswap \qwx[2]             & \qw    & \qw                        & \qw    & \qw            & \qw    & \qswap \qwx[2]             & \qw   & \qw                        & \qw   & \qw            & \qw   & \qw            & \qswap                          & \qw            & \qw   & \qw                & \qw                & \qw       & \qw\\      
	 	\lstick{l: \ket{0}}                 & \qw                   & \qw                              & \qw        & \qw            & \qw   & \qw        & \qw     & \qw                        & \qw    & \qswap \qwx[1]             & \qw    & \qw            & \qw    & \qw                        & \qw   & \qswap \qwx[1]             & \qw   & \qw            & \qw   & \qw            & \qw                             & \qw            & \qw   & \meter             & \qw                & \qw       & \qw\\      
	 	\lstick{m: \ket{0}}                 & \qw                   & \qw                              & \qw        & \qw            & \qw   & \gate{H}   & \qw     & \ctrl{1}                   & \qw    & \ctrl{2}                   & \qw    & \gate{X}       & \qw    & \ctrl{3}                   & \qw   & \ctrl{4}                   & \qw   & \gate{X}       & \qw   & \qw            & \qw                             & \qw            & \qw   & \qw                & \qw                & \qw       & \qw\\      
	 	\lstick{x_1: \ket{0}}               & \gate{H}              & \gate{R^\dagger_z(\pi)}          & \gate{S}   & \qw            & \qw   & \qw        & \qw     & \qswap                     & \qw    & \qw                        & \qw    & \qw            & \qw    & \qw                        & \qw   & \qw                        & \qw   & \qw            & \qw   & \qw            & \qw                             & \qw            & \qw   & \qw                & \qw                & \qw       & \qw\\      
	 	\lstick{l_1: \ket{0}}               & \gate{Id}             & \qw                              & \qw        & \qw            & \qw   & \qw        & \qw     & \qw                        & \qw    & \qswap                     & \qw    & \qw            & \qw    & \qw                        & \qw   & \qw                        & \qw   & \qw            & \qw   & \qw            & \qw                             & \qw            & \qw   & \qw                & \qw                & \qw       & \qw\\      
	 	\lstick{x_2: \ket{0}}               & \gate{H}              & \gate{R^\dagger_z(\pi)}          & \gate{S}   & \gate{Z}       & \qw   & \qw        & \qw     & \qw                        & \qw    & \qw                        & \qw    & \qw            & \qw    & \qswap                     & \qw   & \qw                        & \qw   & \qw            & \qw   & \qw            & \qw                             & \qw            & \qw   & \qw                & \qw                & \qw       & \qw\\      
	 	\lstick{l_2: \ket{0}}               & \gate{X}              & \qw                              & \qw        & \qw            & \qw   & \qw        & \qw     & \qw                        & \qw    & \qw                        & \qw    & \qw            & \qw    & \qw                        & \qw   & \qswap                     & \qw   & \qw            & \qw   & \qw            & \qw                             & \qw            & \qw   & \qw                & \qw                & \qw       & \qw\\      
	 	\lstick{c_{0}: 0}                   & \cw                   & \cw                              & \cw        & \cw            & \cw   & \cw        & \cw     & \cw                        & \cw    & \cw                        & \cw    & \cw            & \cw    & \cw                        & \cw   & \cw                        & \cw   & \cw            & \cw   & \cw            & \cw                             & \cw            & \cw   & \cw                & \cw \cwx[-9]       & \cw       & \cw\\      
	 	\lstick{c_{1}: 0}                   & \cw                   & \cw                              & \cw        & \cw            & \cw   & \cw        & \cw     & \cw                        & \cw    & \cw                        & \cw    & \cw            & \cw    & \cw                        & \cw   & \cw                        & \cw   & \cw            & \cw   & \cw            & \cw                             & \cw            & \cw   & \cw \cwx[-7]       & \cw                & \cw       & \cw\\      
	 }
}

\newcommand{\TenCopiesHighLevelCircuit}{
	\Qcircuit @C=0.5em @R=0.0em @!R {
		\lstick{a_{0}: \ket{0}} & \qw & \qw & \qw & \qw & \qw & \qw & \qw & \qw & \qw & \qw & \qw & \qw & \qw & \qw & \qw & \qw & \qw & \qw & \qw & \qw & \qw & \qw & \qw & \qw & \qw & \qw & \qw & \qw & \qw & \qw & \qw & \gate{H} & \ctrl{2} & \ctrl{3} & \ctrl{4} & \ctrl{5} & \ctrl{6} & \ctrl{7} & \ctrl{8} & \ctrl{9} & \ctrl{10} & \ctrl{11} & \gate{H} & \qw & \qw & \meter & \qw & \qw\\
		\lstick{m_{0}: \ket{0}} & \qw & \qw & \qw & \qw & \qw & \qw & \qw & \qw & \qw & \qw & \qw & \qw & \qw & \qw & \qw & \qw & \qw & \qw & \qw & \qw & \gate{H} & \ctrl{1} & \ctrl{2} & \ctrl{3} & \ctrl{4} & \ctrl{5} & \ctrl{6} & \ctrl{7} & \ctrl{8} & \ctrl{9} & \ctrl{10} & \ctrl{11} & \qw & \qw & \qw & \qw & \qw & \qw & \qw & \qw & \qw & \qw & \qw & \qw & \qw & \qw & \qw & \qw\\
		\lstick{d_{0}: \ket{0}} & \qw & \qw & \qw & \qw & \qw & \qw & \qw & \qw & \qw & \qw & \qw & \qw & \qw & \qw & \qw & \qw & \qw & \qw & \gate{H} & \gate{R_z(-3.1)} & \gate{S} & \control\qw & \qw & \qw & \qw & \qw & \qw & \qw & \qw & \qw & \qw & \qw & \qswap & \qw & \qw & \qw & \qw & \qw & \qw & \qw & \qw & \qw & \qw & \qw & \qw & \qw & \qw & \qw\\
		\lstick{d_{1}: \ket{0}} & \qw & \qw & \qw & \qw & \qw & \qw & \qw & \qw & \qw & \qw & \qw & \qw & \qw & \qw & \qw & \qw & \gate{H} & \gate{R_z(-3.1)} & \gate{S} & \qw & \qw & \qw & \control\qw & \qw & \qw & \qw & \qw & \qw & \qw & \qw & \qw & \qw & \qw & \qswap & \qw & \qw & \qw & \qw & \qw & \qw & \qw & \qw & \qw & \qw & \qw & \qw & \qw & \qw\\
		\lstick{d_{2}: \ket{0}} & \qw & \qw & \qw & \qw & \qw & \qw & \qw & \qw & \qw & \qw & \qw & \qw & \qw & \qw & \gate{H} & \gate{R_z(-3.1)} & \gate{S} & \qw & \qw & \qw & \qw & \qw & \qw & \control\qw & \qw & \qw & \qw & \qw & \qw & \qw & \qw & \qw & \qw & \qw & \qswap & \qw & \qw & \qw & \qw & \qw & \qw & \qw & \qw & \qw & \qw & \qw & \qw & \qw\\
		\lstick{d_{3}: \ket{0}} & \qw & \qw & \qw & \qw & \qw & \qw & \qw & \qw & \qw & \qw & \qw & \qw & \gate{H} & \gate{R_z(-3.1)} & \gate{S} & \qw & \qw & \qw & \qw & \qw & \qw & \qw & \qw & \qw & \control\qw & \qw & \qw & \qw & \qw & \qw & \qw & \qw & \qw & \qw & \qw & \qswap & \qw & \qw & \qw & \qw & \qw & \qw & \qw & \qw & \qw & \qw & \qw & \qw\\
		\lstick{d_{4}: \ket{0}} & \qw & \qw & \qw & \qw & \qw & \qw & \qw & \qw & \qw & \qw & \gate{H} & \gate{R_z(-3.1)} & \gate{S} & \qw & \qw & \qw & \qw & \qw & \qw & \qw & \qw & \qw & \qw & \qw & \qw & \control\qw & \qw & \qw & \qw & \qw & \qw & \qw & \qw & \qw & \qw & \qw & \qswap & \qw & \qw & \qw & \qw & \qw & \qw & \qw & \qw & \qw & \qw & \qw\\
		\lstick{d_{5}: \ket{0}} & \qw & \qw & \qw & \qw & \qw & \qw & \qw & \qw & \gate{H} & \gate{R_z(-3.1)} & \gate{S} & \qw & \qw & \qw & \qw & \qw & \qw & \qw & \qw & \qw & \qw & \qw & \qw & \qw & \qw & \qw & \control\qw & \qw & \qw & \qw & \qw & \qw & \qw & \qw & \qw & \qw & \qw & \qswap & \qw & \qw & \qw & \qw & \qw & \qw & \qw & \qw & \qw & \qw\\
		\lstick{d_{6}: \ket{0}} & \qw & \qw & \qw & \qw & \qw & \qw & \gate{H} & \gate{R_z(-3.1)} & \gate{S} & \qw & \qw & \qw & \qw & \qw & \qw & \qw & \qw & \qw & \qw & \qw & \qw & \qw & \qw & \qw & \qw & \qw & \qw & \control\qw & \qw & \qw & \qw & \qw & \qw & \qw & \qw & \qw & \qw & \qw & \qswap & \qw & \qw & \qw & \qw & \qw & \qw & \qw & \qw & \qw\\
		\lstick{d_{7}: \ket{0}} & \qw & \qw & \qw & \qw & \gate{H} & \gate{R_z(-3.1)} & \gate{S} & \qw & \qw & \qw & \qw & \qw & \qw & \qw & \qw & \qw & \qw & \qw & \qw & \qw & \qw & \qw & \qw & \qw & \qw & \qw & \qw & \qw & \control\qw & \qw & \qw & \qw & \qw & \qw & \qw & \qw & \qw & \qw & \qw & \qswap & \qw & \qw & \qw & \qw & \qw & \qw & \qw & \qw\\
		\lstick{d_{8}: \ket{0}} & \qw & \qw & \gate{H} & \gate{R_z(-3.1)} & \gate{S} & \qw & \qw & \qw & \qw & \qw & \qw & \qw & \qw & \qw & \qw & \qw & \qw & \qw & \qw & \qw & \qw & \qw & \qw & \qw & \qw & \qw & \qw & \qw & \qw & \control\qw & \qw & \qw & \qw & \qw & \qw & \qw & \qw & \qw & \qw & \qw & \qswap & \qw & \qw & \qw & \qw & \qw & \qw & \qw\\
		\lstick{d_{9}: \ket{0}} & \gate{H} & \gate{R_z(-3.1)} & \gate{S} & \qw & \qw & \qw & \qw & \qw & \qw & \qw & \qw & \qw & \qw & \qw & \qw & \qw & \qw & \qw & \qw & \qw & \qw & \qw & \qw & \qw & \qw & \qw & \qw & \qw & \qw & \qw & \control\qw & \qw & \qw & \qw & \qw & \qw & \qw & \qw & \qw & \qw & \qw & \qswap & \qw & \qw & \qw & \qw & \qw & \qw\\
		\lstick{l_{0}: \ket{0}} & \qw & \qw & \qw & \qw & \qw & \qw & \qw & \qw & \qw & \qw & \qw & \qw & \qw & \qw & \qw & \qw & \qw & \qw & \qw & \qw & \qw & \qw & \qw & \qw & \qw & \qw & \qw & \qw & \qw & \qw & \qw & \targ & \qw & \qw & \qw & \qw & \qw & \qw & \qw & \qw & \qw & \qw & \qw & \qw & \meter & \qw & \qw & \qw\\
		\lstick{in_{0}: \ket{0}} & \gate{R_x(3.1)} & \qw & \qw & \qw & \qw & \qw & \qw & \qw & \qw & \qw & \qw & \qw & \qw & \qw & \qw & \qw & \qw & \qw & \qw & \qw & \qw & \qw & \qw & \qw & \qw & \qw & \qw & \qw & \qw & \qw & \qw & \qw & \qswap \qwx[-11] & \qw & \qw & \qw & \qw & \qw & \qw & \qw & \qw & \qw & \qw & \qw & \qw & \qw & \qw & \qw\\
		\lstick{in_{1}: \ket{0}} & \gate{R_x(3.1)} & \qw & \qw & \qw & \qw & \qw & \qw & \qw & \qw & \qw & \qw & \qw & \qw & \qw & \qw & \qw & \qw & \qw & \qw & \qw & \qw & \qw & \qw & \qw & \qw & \qw & \qw & \qw & \qw & \qw & \qw & \qw & \qw & \qswap \qwx[-11] & \qw & \qw & \qw & \qw & \qw & \qw & \qw & \qw & \qw & \qw & \qw & \qw & \qw & \qw\\
		\lstick{in_{2}: \ket{0}} & \gate{R_x(3.1)} & \qw & \qw & \qw & \qw & \qw & \qw & \qw & \qw & \qw & \qw & \qw & \qw & \qw & \qw & \qw & \qw & \qw & \qw & \qw & \qw & \qw & \qw & \qw & \qw & \qw & \qw & \qw & \qw & \qw & \qw & \qw & \qw & \qw & \qswap \qwx[-11] & \qw & \qw & \qw & \qw & \qw & \qw & \qw & \qw & \qw & \qw & \qw & \qw & \qw\\
		\lstick{in_{3}: \ket{0}} & \gate{R_x(3.1)} & \qw & \qw & \qw & \qw & \qw & \qw & \qw & \qw & \qw & \qw & \qw & \qw & \qw & \qw & \qw & \qw & \qw & \qw & \qw & \qw & \qw & \qw & \qw & \qw & \qw & \qw & \qw & \qw & \qw & \qw & \qw & \qw & \qw & \qw & \qswap \qwx[-11] & \qw & \qw & \qw & \qw & \qw & \qw & \qw & \qw & \qw & \qw & \qw & \qw\\
		\lstick{in_{4}: \ket{0}} & \gate{R_x(3.1)} & \qw & \qw & \qw & \qw & \qw & \qw & \qw & \qw & \qw & \qw & \qw & \qw & \qw & \qw & \qw & \qw & \qw & \qw & \qw & \qw & \qw & \qw & \qw & \qw & \qw & \qw & \qw & \qw & \qw & \qw & \qw & \qw & \qw & \qw & \qw & \qswap \qwx[-11] & \qw & \qw & \qw & \qw & \qw & \qw & \qw & \qw & \qw & \qw & \qw\\
		\lstick{in_{5}: \ket{0}} & \gate{R_x(3.1)} & \qw & \qw & \qw & \qw & \qw & \qw & \qw & \qw & \qw & \qw & \qw & \qw & \qw & \qw & \qw & \qw & \qw & \qw & \qw & \qw & \qw & \qw & \qw & \qw & \qw & \qw & \qw & \qw & \qw & \qw & \qw & \qw & \qw & \qw & \qw & \qw & \qswap \qwx[-11] & \qw & \qw & \qw & \qw & \qw & \qw & \qw & \qw & \qw & \qw\\
		\lstick{in_{6}: \ket{0}} & \gate{R_x(3.1)} & \qw & \qw & \qw & \qw & \qw & \qw & \qw & \qw & \qw & \qw & \qw & \qw & \qw & \qw & \qw & \qw & \qw & \qw & \qw & \qw & \qw & \qw & \qw & \qw & \qw & \qw & \qw & \qw & \qw & \qw & \qw & \qw & \qw & \qw & \qw & \qw & \qw & \qswap \qwx[-11] & \qw & \qw & \qw & \qw & \qw & \qw & \qw & \qw & \qw\\
		\lstick{in_{7}: \ket{0}} & \gate{R_x(3.1)} & \qw & \qw & \qw & \qw & \qw & \qw & \qw & \qw & \qw & \qw & \qw & \qw & \qw & \qw & \qw & \qw & \qw & \qw & \qw & \qw & \qw & \qw & \qw & \qw & \qw & \qw & \qw & \qw & \qw & \qw & \qw & \qw & \qw & \qw & \qw & \qw & \qw & \qw & \qswap \qwx[-11] & \qw & \qw & \qw & \qw & \qw & \qw & \qw & \qw\\
		\lstick{in_{8}: \ket{0}} & \gate{R_x(3.1)} & \qw & \qw & \qw & \qw & \qw & \qw & \qw & \qw & \qw & \qw & \qw & \qw & \qw & \qw & \qw & \qw & \qw & \qw & \qw & \qw & \qw & \qw & \qw & \qw & \qw & \qw & \qw & \qw & \qw & \qw & \qw & \qw & \qw & \qw & \qw & \qw & \qw & \qw & \qw & \qswap \qwx[-11] & \qw & \qw & \qw & \qw & \qw & \qw & \qw\\
		\lstick{in_{9}: \ket{0}} & \gate{R_x(3.1)} & \qw & \qw & \qw & \qw & \qw & \qw & \qw & \qw & \qw & \qw & \qw & \qw & \qw & \qw & \qw & \qw & \qw & \qw & \qw & \qw & \qw & \qw & \qw & \qw & \qw & \qw & \qw & \qw & \qw & \qw & \qw & \qw & \qw & \qw & \qw & \qw & \qw & \qw & \qw & \qw & \qswap \qwx[-11] & \qw & \qw & \qw & \qw & \qw & \qw\\
		\lstick{c_{0}: 0} & \cw & \cw & \cw & \cw & \cw & \cw & \cw & \cw & \cw & \cw & \cw & \cw & \cw & \cw & \cw & \cw & \cw & \cw & \cw & \cw & \cw & \cw & \cw & \cw & \cw & \cw & \cw & \cw & \cw & \cw & \cw & \cw & \cw & \cw & \cw & \cw & \cw & \cw & \cw & \cw & \cw & \cw & \cw & \cw & \cw & \cw \cwx[-23] & \cw & \cw\\
		\lstick{c_{1}: 0} & \cw & \cw & \cw & \cw & \cw & \cw & \cw & \cw & \cw & \cw & \cw & \cw & \cw & \cw & \cw & \cw & \cw & \cw & \cw & \cw & \cw & \cw & \cw & \cw & \cw & \cw & \cw & \cw & \cw & \cw & \cw & \cw & \cw & \cw & \cw & \cw & \cw & \cw & \cw & \cw & \cw & \cw & \cw & \cw & \cw \cwx[-12] & \cw & \cw & \cw\\
	}
}